%% file: main.tex
\documentclass[sigplan,10pt]{acmart}

\acmConference[PPoPP'27]{ACM SIGPLAN Annual Symposium on Principles and Practice of Parallel Programming}{March 2027}{Salt Lake City, UT}
\setcopyright{acmcopyright}
\copyrightyear{2018}
\acmYear{2018}
\acmDOI{XXXXXXX.XXXXXXX}

\usepackage{booktabs}   
\usepackage{subcaption} 
\usepackage{caption}
\usepackage{colortbl}
\usepackage{bigstrut}
\usepackage{microtype}
\usepackage{tabularx}
\usepackage{multirow}
\usepackage{multicol}
\usepackage{rotating}
\usepackage{lipsum}
\usepackage{balance}
\usepackage{mfirstuc}
\usepackage{titlecaps}
\usepackage{listings}
\usepackage{float}
\usepackage[rightcaption]{sidecap}
\usepackage{xr}
\usepackage{xcolor}

\input{sty.tex}

\begin{document}

\title[Toggle Tree for Parallel Graph Processing]{\LARGE Toggle Tree: A New Data Structure for Efficient Parallel Graph Processing}

\author{Xinjian Jiang}
\email{xjian140@ucr.edu}
\affiliation{%
  \institution{University of California, Riverside}
  \city{Riverside}
  \country{USA}
}

\author{Youzhe Liu}
\email{yliu908@ucr.edu}
\affiliation{%
  \institution{University of California, Riverside}
  \city{Riverside}
  \country{USA}
}

\author{Yan Gu}
\email{ygu@cs.ucr.edu}
\affiliation{%
  \institution{University of California, Riverside}
  \city{Riverside}
  \country{USA}
}

\author{Yihan Sun}
\email{yihans@cs.ucr.edu}
\affiliation{%
  \institution{University of California, Riverside}
  \city{Riverside}
  \country{USA}
}

\input{main/00_abstract}


\maketitle

\input{macro.tex}
\input{main/01_introduction.tex}

\input{main/02_preliminary.tex}
\input{main/03_algorithm.tex}

\input{main/04_application.tex}

\input{main/05_experiment.tex}

\input{main/06_relatedwork.tex}
\input{main/07_conclusion.tex}

\bibliographystyle{ACM-Reference-Format}
\balance
\bibliography{bib/strings, bib/main}

\clearpage\onecolumn
\input{main/99_appendix.tex}

\end{document}

%% file: sty.tex
\def\conference
\usepackage{etoolbox}
\newcommand{\ifconference}[1]{{{\ifx\fullversion\undefined{#1}\fi}\xspace}}
\newcommand{\iffullversion}[1]{{{\ifx\conference\undefined{#1}\fi}\xspace}}

\usepackage{graphicx}  
\usepackage{lipsum}  
\newcommand{\hide}[1]{} 
\usepackage{xspace}
\usepackage{textcomp}
\usepackage{comment} 
\usepackage{verbatim}
\usepackage{changepage}

\usepackage{url}

\newcommand{\defn}[1]{\emph{\textbf{#1}}} 
\newcommand{\emp}[1]{\emph{\textbf{#1}}} 
\newcommand{\fname}[1]{\textsc{#1}} 
\newcommand{\vname}[1]{\mathit{#1}} 
\newcommand{\mathfunc}[1]{\mathit{#1}}

\usepackage[ruled,lined,linesnumbered,noend]{algorithm2e}
\usepackage[noend]{algpseudocode}

\makeatletter
\patchcmd{\@algocf@start}
  {-1.5em}
  {0pt}
  {}{}
\newcommand{\nosemic}{\renewcommand{\@endalgocfline}{\relax}}
\newcommand{\dosemic}{\renewcommand{\@endalgocfline}{\algocf@endline}}

\SetSideCommentLeft
\SetKwInput{notations}{Notations}
\SetKwInput{notes}{Notes}
\SetKwInput{maintains}{Maintains}

\SetKwProg{myfunc}{Function}{}{}
\SetKwFor{parForEach}{ParallelForEach}{do}{endfor}
\SetKwFor{Justrepeat}{Repeat}{}{}
\SetKwFor{lrepeat}{repeat until}{}{}

\SetKw{MIN}{min}
\SetKw{MAX}{max}
\SetKw{OR}{or}
\SetKw{AND}{and}

\SetCommentSty{mycommfont}

\SetAlCapFnt{\small\bfseries} 

\SetAlCapNameFnt{\small}

\usepackage{cleveref}
\crefname{section}{Sec.}{Sec.}
\crefname{theorem}{Thm.}{Thm.}
\crefname{lemma}{Lem.}{Lem.}
\crefname{corollary}{Cor.}{Cor.}
\crefname{table}{Tab.}{Tab.}
\crefname{algorithm}{Alg.}{Alg.}
\crefname{figure}{Fig.}{Fig.}
\crefname{fact}{Fact}{Fact}
\Crefname{table}{Tab.}{Tab.}
\crefname{problem}{Problem}{Problem}

\newtheorem{theorem}{Theorem}[section]

\newtheorem{corollary}[theorem]{Corollary}

\newtheorem{fact}[theorem]{Fact}

\let \originalleft \left
\let\originalright\right
\renewcommand{\left}{\mathopen{}\mathclose\bgroup\originalleft}
\renewcommand{\right}{\aftergroup\egroup\originalright}

\usepackage{scalerel} 

\newtheoremstyle{exampstyle}
{.5em} 
{1em} 
{\it} 
{.5em} 
{\it \bfseries} 
{.} 
{.5em} 
{} 
\theoremstyle{exampstyle} 
\theoremstyle{exampstyle} 
\theoremstyle{exampstyle} 
\theoremstyle{exampstyle}

\newcommand{\whp}[1]{\emph{whp}}

\DeclareMathOperator*{\polylog}{polylog}

\newcommand{\true}{\emph{true}}
\newcommand{\false}{\emph{false}}

\usepackage{pifont}

\newcommand{\modelop}[1]{\texttt{#1}}
\newcommand{\forkins}{\modelop{fork}}

\newcommand{\thread}{thread}

\newcommand{\insformat}[1]{\texttt{#1}}

\newcommand{\WriteMin}{\insformat{write\_min}\xspace}

\newcommand{\writemin}{\WriteMin}

\usepackage[shortlabels]{enumitem}

\usepackage{float}
\usepackage[labelfont=bf,font={small}]{caption}

\usepackage[labelfont=bf,list=true]{subcaption}
\usepackage[rightcaption]{sidecap}

\usepackage{wrapfig}

\usepackage{array}
\newcolumntype{L}[1]{>{\raggedright\let\newline\\\arraybackslash\hspace{0pt}}m{#1}}
\newcolumntype{C}[1]{>{\centering\let\newline\\\arraybackslash\hspace{0pt}}m{#1}}
\newcolumntype{R}[1]{>{\raggedleft\let\newline\\\arraybackslash\hspace{0pt}}m{#1}}
\newcolumntype{B}{>{\bf}c}
\usepackage{rotating}

\usepackage{booktabs} 
\usepackage{multicol,multirow}
\usepackage{longtable} 
\usepackage{supertabular} 
\usepackage{colortbl}
\usepackage{bigstrut}

\newcommand{\myparagraph}[1]{\boldmath\vspace{.1em}\noindent\emp{#1}\enspace\unboldmath}

\usepackage{mdframed}
\definecolor{framelinecolor}{RGB}{68,114,196}
\mdfdefinestyle{mystyle}{linecolor=framelinecolor,innertopmargin=1pt,innerbottommargin=2pt,backgroundcolor=gray!20,skipabove=2pt,skipbelow=0pt}
\mdfdefinestyle{densestyle}{linecolor=framelinecolor,innertopmargin=0,innerbottommargin=0,leftmargin=0,rightmargin=0,backgroundcolor=gray!20}
\mdfdefinestyle{compactcode}{linecolor=framelinecolor,innertopmargin=1pt,innerbottommargin=1pt,backgroundcolor=gray!20,skipabove=0pt,skipbelow=0pt,leftmargin=0,rightmargin=0}

\usepackage{framed}

\usepackage{listings}
\newcommand{\codeskip}{{\vspace{.04in}}}

\newdimen\zzsize
\newdimen\kwsize
\newcommand{\basicstyle}{\fontsize{\zzsize}{1\zzsize}\ttfamily}
\newcommand{\keywordstyle}{\fontsize{\kwsize}{1\kwsize}\ttfamily\bf}

\newdimen\zzlstwidth
\usepackage{tikz} 

\setcopyright{none}
\renewcommand\footnotetextcopyrightpermission[1]{} 

%% file: main/00_abstract.tex
\begin{abstract}
  
We present the \textsf{Toggle Tree}\xspace, a parallel data structure for frontier-based graph algorithms. 
A \textsf{Toggle Tree}\xspace is used to represent a subset of vertices using hierarchical bit vectors, 
enabling efficient parallel updates, 
traversal, and set-wise reductions without global packing. 
It provides an \textsf{IndexSet}\xspace interface that unifies a wide range of parallel graph algorithms under a common abstraction of frontiers and active sets, 
and an \textsf{IndexMap}\xspace interface that also incorporates priority-queue semantics. 
Specifically, the \textsf{Toggle Tree}\xspace aims to combine the advantages of traditional sparse and dense frontier representations, 
offering both provable theoretical guarantees and high efficiency in practice.

Using \textsf{Toggle Tree}\xspace{s}, we implement five fundamental graph algorithms, 
including breadth-first search (BFS), $k$-core\xspace, degree-based graph coloring, single-source shortest paths using Bellman-Ford, and weighted-BFS (wBFS). 
Across a diverse set of large real-world graphs and the five graph problems, our implementations outperform state-of-the-art baselines, 
including GBBS and PASGAL, achieving substantial speedups. Our code is publicly available.
\end{abstract}

%% file: macro.tex
\providecommand{\degree}{d}
\newcommand{\dmin}{k}
\newcommand{\dbar}{d_{\text{\it avg}}}
\newcommand{\nei}{N}
\newcommand{\degreestar}{\tilde{d}}
\newcommand{\induceddegree}{induced degree}
\newcommand{\kmax}{k_{\max}}
\newcommand{\numpeeling}{\mathfunc{\rho}\xspace}
\newcommand{\kcore}{$k$-core\xspace}
\newcommand{\coloring}{coloring\xspace}
\newcommand{\atominc}{\fname{atomic\_inc}}
\newcommand{\atomdec}{\fname{atomic\_dec}}
\newcommand{\subround}{subround\xspace}
\newcommand{\HBS}{HBS\xspace}

\newcommand{\bagarray}{\vname{\mathcal{B}}\xspace}
\newcommand{\bagput}{\mf{BagInsert}\xspace}
\newcommand{\bagforall}{\mf{BagForAll}\xspace}
\newcommand{\bagpack}{\mf{BagExtractAll}\xspace}
\newcommand{\bagfind}{\fname{BagFind}\xspace}
\newcommand{\bagresize}{\fname{try\_resize}\xspace}
\newcommand{\coreness}{\mbox{{coreness}}\xspace}

\newcommand{\mf}[1]{{\text{\scshape#1}}}

\newcommand{\stride}{\vname{stride}\xspace}
\newcommand{\truecoreness}{{\kappa}\xspace}
\newcommand{\maxcoreness}{{k_{\max}}\xspace}
\newcommand{\finishsign}{\vname{finished}\xspace}
\newcommand{\frontier}{{\mathcal{F}}\xspace}
\newcommand{\nextfrontier}{\mathcal{F}_{\text{next}}}
\newcommand{\alive}{active}
\newcommand{\alivevertices}{{\mathcal{A}}\xspace}
\newcommand{\samplemode}{\vname{sample\_mode}\xspace}
\newcommand{\samplecond}{\vname{sample\_cond}\xspace}
\newcommand{\sampleprob}{\vname{sample\_prob}\xspace}
\newcommand{\errorp}{\vname{error\_prob}\xspace}
\newcommand{\samplerate}{\vname{sample\_rate}\xspace}
\newcommand{\oldV}{\vname{oldV}\xspace}
\newcommand{\newV}{\vname{newV}\xspace}

\newcommand{\FMain}{\mf{Coreness}\xspace}
\newcommand{\FAddToBucket}{\mf{AddToBucket}\xspace}
\newcommand{\FPeel}{\mf{Peel}\xspace}
\newcommand{\FProcessBucket}{\mf{SolveAndUpdate}}
\newcommand{\FProcessNeighbors}{\mf{ProcessNeighbors}\xspace}
\newcommand{\FError}{\mf{Validate}\xspace}
\newcommand{\FInitSampler}{\mf{SetSampler}\xspace}
\newcommand{\FSetSampler}{\mf{Resample}\xspace}
\newcommand{\FCheckSampleSecurity}{\mf{CheckSample}\xspace}
\newcommand{\FCountVertex}{\mf{DegreeCount}\xspace}
\newcommand{\FMapNeighborsParallel}{\mf{MapNeighborsParallel}\xspace}
\newcommand{\FMapNeighborsSequential}{\mf{MapNeighborsSequential}\xspace}
\newcommand{\FFetchAndAddVertex}{\mf{FetchAndAddVertex}\xspace}
\newcommand{\FSampleVertex}{\mf{AddSample}\xspace}
\newcommand{\FExpDec}{\mf{ExpDec}\xspace}
\newcommand{\FFilterOut}{\mf{FilterOut}\xspace}
\newcommand{\FExtractAll}{\mf{ExtractAll}()\xspace}
\newcommand{\FCalcerror}{\mf{CalcError}\xspace}
\newcommand{\FHistogram}{\mf{Histogram}\xspace}
\newcommand{\pack}{\mf{Pack}}
\newcommand{\FBuildBuckets}[1]{\mf{BuildBuckets(}#1\mf{)}\xspace}
\newcommand{\FGetNextBucket}{\mf{GetNextBucket()}\xspace}
\newcommand{\FUpdate}[1]{\mf{DecreaseKey(}#1\mf{)}\xspace}
\newcommand{\FFilter}{\mf{Filter}\xspace}
\newcommand{\FGetBucket}{\mf{GetBucket}\xspace}
\newcommand{\Fdecdeg}{\mf{DecDegree}\xspace}
\newcommand{\Fonlinepeelwithsampling}{\mf{OnlinePeelWithSampling}\xspace}

\newcommand{\customIf}[2]{%
    \textbf{if} #1 \textbf{then} #2 
}
\newcommand{\customForEach}[2]{%
    \textbf{foreach} #1 \textbf{do} #2
}

\newcommand{\exphits}{\mathfunc{\mu}\xspace}
\newcommand{\samplerateproof}{\mathfunc{p}\xspace}
\newcommand{\reducerate}{r}
\newcommand{\sampler}{\sigma}
\newcommand{\countingbag}{\mathcal{C}}

\newcommand{\mode}{\text{\it mode}}
\newcommand{\rate}{\text{\it rate}}
\newcommand{\cnt}{\text{\it cnt}}
\newcommand{\cntproof}{s}
\newcommand{\rateproof}{p}

\newcommand{\PKC}{\textsf{PKC}\xspace}
\newcommand{\pkc}{\PKC}
\newcommand{\ParK}{\textsf{ParK}\xspace}
\newcommand{\Park}{\ParK}
\newcommand{\park}{\ParK}
\newcommand{\Julienne}{\textsf{Julienne}\xspace}
\newcommand{\julienne}{\Julienne{}}
\newcommand{\GBBS}{\textsf{Julienne}\xspace}
\newcommand{\BZ}{\textsf{BZ}\xspace}

\newcommand{\LJ}{\emph{LJ}\xspace}
\newcommand{\FB}{\emph{FB}\xspace}
\newcommand{\OK}{\emph{OK}\xspace}
\newcommand{\WB}{\emph{WB}\xspace}
\newcommand{\TW}{\emph{TW}\xspace}
\newcommand{\FS}{\emph{FS}\xspace}
\newcommand{\WK}{\emph{WK}\xspace}
\newcommand{\EW}{\emph{EW}\xspace}
\newcommand{\EH}{\emph{EH}\xspace}
\newcommand{\SD}{\emph{SD}\xspace}
\newcommand{\CW}{\emph{CW}\xspace}
\newcommand{\HL}{\emph{HL14}\xspace}
\newcommand{\HLs}{\emph{HL12}\xspace}
\newcommand{\AF}{\emph{AF}\xspace}
\newcommand{\NA}{\emph{NA}\xspace}
\newcommand{\AS}{\emph{AS}\xspace}
\newcommand{\EU}{\emph{EU}\xspace}
\newcommand{\CH}{\emph{CH}\xspace}
\newcommand{\GLt}{\emph{GL2}\xspace}
\newcommand{\GL}{\emph{GL5}\xspace}
\newcommand{\GLs}{\emph{GL}\xspace}
\newcommand{\COS}{\emph{COS}\xspace}
\newcommand{\TRCE}{\emph{TRCE}\xspace}
\newcommand{\BBL}{\emph{BBL}\xspace}
\newcommand{\GRD}{\emph{GRID}\xspace}
\newcommand{\CBC}{\emph{CUBE}\xspace}
\newcommand{\HCNS}{\emph{HCNS}\xspace}
\newcommand{\HPL}{\emph{HPL}\xspace}
\newcommand{\knn}{$k$-NN}
\newcommand{\IC}{\emph{IC}\xspace}

\newcommand{\LJfull}{\emph{soc-LiveJournal1}\xspace}
\newcommand{\FBfull}{\emph{socfb-konect }\xspace}
\newcommand{\OKfull}{\emph{com-orkut}\xspace}
\newcommand{\WBfull}{\emph{soc-sinaweibo}\xspace}
\newcommand{\TWfull}{\emph{Twitter}\xspace}
\newcommand{\FSfull}{\emph{Friendster}\xspace}
\newcommand{\WKfull}{\emph{enwiki-2023}\xspace}
\newcommand{\EHfull}{\emph{eu-host}\xspace}
\newcommand{\SDfull}{\emph{sd-arc}\xspace}
\newcommand{\CWfull}{\emph{ClueWeb}\xspace}
\newcommand{\HLfull}{\emph{Hyperlink14}\xspace}
\newcommand{\HLsfull}{\emph{Hyperlink12}\xspace}
\newcommand{\AFfull}{\emph{OSM Africa}\xspace}
\newcommand{\NAfull}{\emph{OSM North America}\xspace}
\newcommand{\ASfull}{\emph{OSM Asia}\xspace}
\newcommand{\EUfull}{\emph{OSM Europe}\xspace}
\newcommand{\CHfull}{\emph{Chem}\xspace}
\newcommand{\GLtfull}{\emph{GeoLife2}\xspace}
\newcommand{\GLfull}{\emph{GeoLife5}\xspace}
\newcommand{\GLsfull}{\emph{GeoLife10}\xspace}
\newcommand{\COSfull}{\emph{Cosmo50}\xspace}
\newcommand{\TRCEfull}{\emph{Huge Traces}\xspace}
\newcommand{\BBLfull}{\emph{Huge Bubbles}\xspace}
\newcommand{\GRDfull}{\emph{Huge Grids}\xspace}
\newcommand{\CBCfull}{\emph{Huge Cubic}\xspace}
\newcommand{\HCNSfull}{\emph{Huge Coreness}\xspace}
\newcommand{\HPLfull}{\emph{Huge Power Law}\xspace}
\newcommand{\ICfull}{\emph{Indochina}}

\newcommand{\KNN}{$k$-NN\xspace}

\newcommand{\yan}[1]{{\color{violet}{\bf Yan:} #1}}
\newcommand{\yihan}[1]{{\color{blue}{\bf Yihan:} #1}}
\newcommand{\xiaojun}[1]{{\color{cyan}{\bf Xiaojun:} #1}}
\newcommand{\youzhe}[1]{{\color{cyan}{\bf Youzhe:} #1}}
\newcommand{\xinjian}[1]{{\color{cyan}{\bf Xinjian:} #1}}
\newcommand{\fillin}[1]{{\color{red} #1}} 

\newcommand{\revise}[1]{{\color{black} #1}}

\newcommand{\VS}{vertex subset\xspace}
\newcommand{\DataStructure}{\textsf{Toggle Tree}\xspace}
\newcommand{\ourtree}{\DataStructure}
\newcommand{\DS}{\textsf{ours}\xspace}
\newcommand{\DataTypeA}{\textsf{IndexSet}\xspace}
\newcommand{\DataTypeB}{\textsf{IndexMap}\xspace}
\newcommand{\DataTypeIS}{\textsf{IndexSet}\xspace}
\newcommand{\DataTypeIM}{\textsf{IndexMap}\xspace}
\newcommand{\hashbag}{hash bag\xspace} 

\newcommand{\ff}{\mathcal{F}}
\newcommand{\fa}{\mathcal{A}}
\newcommand{\ffset}{\mathcal{F}_{\mathsf{set}}}
\newcommand{\ffelm}{\mathcal{F}_{\mathsf{elm}}}
\newcommand{\faset}{\mathcal{A}_{\mathsf{set}}}
\newcommand{\faelm}{\mathcal{A}_{\mathsf{elm}}}
\newcommand{\fafro}{\mathcal{A}_{\mathsf{fro}}}

\newcommand{\myfuncname}[1]{\textsf{#1}}

\newcommand{\finsert}{\myfuncname{Insert}}
\newcommand{\fdelete}{\myfuncname{Remove}}
\newcommand{\fremove}{\myfuncname{Remove}}
\newcommand{\finit}{\myfuncname{Init}}
\newcommand{\fempty}{\myfuncname{Empty}}
\newcommand{\fforeach}{\myfuncname{ForEach}}
\newcommand{\fcontains}{\myfuncname{Contains}}
\newcommand{\fclear}{\myfuncname{Clear}} 
\newcommand{\fgetval}{\myfuncname{GetVal}} 
\newcommand{\fdecval}{\myfuncname{DecVal}}
\newcommand{\fextract}{\myfuncname{ExtractMin}}
\newcommand{\freduce}{\myfuncname{Reduce}}
\newcommand{\frepair}{\myfuncname{Repair}}
\newcommand{\val}{v^*}
\newcommand{\bits}{\vname{toggle}} 
\newcommand{\ind}{\vname{index}}
\newcommand{\faor}{{\insformat{fetch\_and\_or}}\xspace}
\newcommand{\faand}{{\insformat{fetch\_and\_and}}\xspace}
\newcommand{\faadec}{{\insformat{fetch\_and\_sub}}\xspace}
\newcommand{\fetchandor}{\faor}
\newcommand{\fetchandand}{\faand}
\newcommand{\cond}{\vname{cond}}
\newcommand{\bitflag}{\vname{flag}}
\newcommand{\propagate}{\vname{toggled}}
\newcommand{\dirty}{\vname{dirty}}
\newcommand{\augval}{\vname{val}}
\newcommand{\minval}{\vname{minval}}

\newcommand{\source}{\vname{source}}

\newcommand{\markatomic}[1]{\hl{#1}}

\newcommand{\setatomic}{{\insformat{atomic\_set}}\xspace}

\newcommand{\popcount}{{\insformat{popcount}}}
\newcommand{\ctz}{{\insformat{ctz}}}
\newcommand{\pdep}{{\insformat{pdep}}}

\newcommand{\Push}{{\textsf{Push}}\xspace}
\newcommand{\Pull}{{\textsf{Pull}}\xspace}
\newcommand{\DependencyExist}{{\textsf{Existential Dependency}}\xspace}
\newcommand{\DependencyAll}{{\textsf{Universal Dependency}}\xspace}
\newcommand{\DependencyCounter}{{\textsf{Global-Threshold Dependency}}\xspace}

\newcommand{\checknumber}[1]{{\color{red}{#1}}}

%% file: main/01_introduction.tex
\section{Introduction}

With growing graph scales, analyzing and processing a graph in parallel has become increasingly important. 
In such tasks, a critical and unavoidable step is to maintain a subset of vertices. 
For instance, many algorithms maintain a \emph{frontier} set as vertices to be processed in a given round. 
Depending on different algorithmic techniques, other dynamic \VS{s} may be needed, such as active set (unprocessed vertices) or candidate sets (vertices ready for processing). 

Efficiently maintaining such a \VS is often critical to algorithm performance. 
The minimum requirements for a \VS{} include dynamic updates via insertions and deletions, and a \emph{for-each} operation to process all vertices in it. 
Some algorithms require operations beyond basic set semantics, such as a priority queue interface. 
Simple as it sounds, maintaining a \VS is highly non-trivial in parallel. 
Due to inherent graph irregularity, these \VS{}s are highly dynamic, with their sizes fluctuating drastically during execution. 
For updates, the \VS{} must handle rapid, \emph{concurrent} modifications, ideally in near-constant time per operation. 
For the \emph{for-each} operation, the \VS{} must remain efficient across all scales---from a handful of vertices to nearly all vertices---ideally with access time proportional to the subset size.  
Sequentially, simple array-based structures like FIFO queues (for vertex tracking) and binary heaps (for priority queries) suffice. 
However, directly using them in parallel incurs significant overhead due to heavy contention (e.g., on queue ends or tree roots).

Existing parallel \VS{}s fall into two categories---the \emph{sparse} form and the \emph{dense} form---and each satisfies one of the two goals above at the expense of the other.
The sparse form stores all vertices explicitly in an array~\cite{shun2013ligra,gbbs2021,dhulipala2020semi,wang2023parallel,dong2024pasgal,dong2021efficient},
supporting a straightforward \emph{for-each} operation.
However, for parallel updates, 
they may either use another round of edge scanning to insert elements synchronously~\cite{shun2013ligra,gbbs2021,dhulipala2020semi}, 
leave empty slots to enable concurrent insertions~\cite{wang2023parallel,dong2024pasgal,dong2021efficient}, 
or use per-thread queues and combine at the end~\cite{beamer2015gap}, 
each incurring additional time and/or space overhead (we introduce more details in \cref{sec:discussion}).
An alternative is the \emph{dense} representation, which simply stores $n$ boolean flags that indicate vertex membership. 
Updates now reduce to a single flag flipping, but a \emph{for-each} must scan the entire array, costing $\Theta(n)$ work regardless of size.
The dense form therefore pays off only when the size is $\Theta(n)$. 
Alongside the structural switching overhead, this limits the benefit of the dense mode to only a few rounds, or entirely precludes it if the frontier remains moderate throughout the execution.

In this paper, we overcome these challenges by proposing a simple yet highly efficient data structure: the \emph{\ourtree}. 
It combines the advantages of the sparse and dense modes. 
The \emph{\ourtree} uses hierarchical bitmasks.
Each vertex is represented by one membership bit (as in the dense mode) at the leaf level.
Each bit in the upper level represents the aggregation (logical OR) of $w$ bits in the lower level, where $w$ is the word width. 
When traversing the set (a \emph{for-each}), empty subtrees can be skipped, with cost $\tilde{O}(s)$ for subset size $s$ (close to the efficiency in the sparse mode). 
For updates, concurrent updates can be applied by simple atomic bit operations, enabling high space efficiency ($\Theta(n)$ bits or $\Theta(n/w)$ words in total for a \ourtree).
As a stand-alone data structure, the \ourtree also eliminates the need to switch between the sparse and dense representations.

For generality, we refer to the abstract data type as an \defn{\DataTypeA{}}, since its interface broadly applies to a subset of a universe of $n$ indices, denoted by integers from $0$ to $n-1$. 
We can also augment the \ourtree{} to extend an \DataTypeA{} to an \DataTypeB{}, where each index is associated with a value, and to support a priority-queue-like interface. 
In particular, we introduce an auxiliary tournament tree that maintains subtree minima under concurrent decrease-key updates and supports batched extraction of the current minimum, 
providing a simple and efficient alternative to parallel priority queues for weighted graph traversals.

For the most typical usage of \VS{}s, where the algorithm processes a \emph{frontier} of vertices in each round, 
we theoretically show that if a parallel graph algorithm finishes in $\rho$ rounds, the amortized cost per insertion, deletion, 
or traversal step in \DataStructure{} is bounded by $O(\frac{\log \rho}{\log\log n})$, assuming all vertices are touched. 
Even in the worst case when $\rho=n$, the overhead is small; however, this indicates that only one vertex is processed per round with no parallelism. 
Usually, it is preferable to run these algorithms in parallel when $\rho=\polylog(n)$, and in these cases, \ourtree{} operations are asymptotically optimal.

In practice, we implement five fundamental parallel graph algorithms using \textit{\ourtree{s}} to demonstrate their efficiency and usability, including BFS, single-source shortest paths (SSSP) using Bellman-Ford, \kcore{}, degree-based graph coloring, and weighted-BFS (wBFS). 
For all of them, we use existing algorithmic techniques and only replace the \VS{} with \ourtree{s}. 
Compared with a collection of state-of-the-art baselines~\cite{gbbs2021,dong2024pasgal,beamer2015gap,nguyen2013lightweight,dasari2014park,kabir2017parallel}, our algorithms are on average 1.8$\times$ faster than the best baseline.
Across 100 tested algorithm-graph instances, our solution outperforms the strongest baseline in 98 of them, achieving at least a 2$\times$ speedup in over one third of the test instances.

Our code is publicly available at~\cite{totcode}.

%% file: main/02_preliminary.tex
\section{Preliminaries}\label{sec:prelim}

\myparagraph{Parallel Computational Model.}
We use the work-span model for classical multi-threaded binary fork-join~\cite{BL98,blelloch2020optimal}.
We assume a set of \thread{}s sharing a common memory.
A process can \forkins{} two child \thread{}s to execute in parallel.
When both children complete, the parent process continues.
The \defn{work} of an algorithm is the total number of instructions.
The \defn{span} of an algorithm is the length of the longest dependency chain among operations.
A computation with work $W$ and span $S$ on $P$ processors can be executed by a randomized work-stealing scheduler~\cite{arora2001thread,BL98,gu2022analysis} in time $W/P + O(S)$ with high probability in $W$. 
We also use hardware-supported atomic primitives, including atomic bitwise operations (\faor{} and \faand{}) for lock-free updates (\cref{sec:operations}).

The analysis of the graph applications using the \ourtree{} (in \cref{sec:applications}) assumes additional unit-cost atomic operations, including $\writemin$ for distance relaxation, and atomic counter decrements in $k$-core and coloring.

\myparagraph{Notations.}
We consider a graph $G = (V, E)$ stored in compressed sparse row (CSR) format, with $n = |V|$ vertices and $m = |E|$ edges (number of directed CSR edge entries).
For a vertex $v$, $N(v)$ denotes its neighbor set and $\degree(v) = |N(v)|$ its degree.
We use $w$ to denote the machine word size (typically 64).
In asymptotic notation, we use $\log_a b$ as shorthand for $1+\log_a{(a+b)}$ for simplicity, so that the bounds remain well-defined in corner cases.

\myparagraph{Frontier-Based Graph Algorithms.}
Many parallel graph algorithms follow a \emph{frontier-based} paradigm.
The algorithm maintains an \emph{active set} $\fa$ of vertices that have not been finalized
and a \emph{frontier} $\ff$ of vertices that are processed in the current round.
Each round of the algorithm processes all $v\in \ff$ (e.g., relaxing outgoing edges),
which may generate the next frontier $\ff'$ and remove processed vertices from $\fa$.
The algorithm terminates when the frontier or the active set becomes empty. 
We use examples of BFS, Bellman-Ford, $k$-core, graph coloring, weighted-BFS, all discussed in \cref{sec:applications}.

%% file: main/03_algorithm.tex
\section{The \ourtree{} Design}\label{sec:algo}

In this section, we describe the design of the \ourtree, which implements the interfaces for \DataTypeA{} and \DataTypeB. 
\DataTypeA{} and \DataTypeB{} are similar to standard sets and maps, except that keys are integers from 0 to $n-1$.
Well-known data structures exist for these data types~\cite{van1977preserving,gu2023parallel,willard1983log}.
In contrast to them, the primary goal of the \ourtree is to achieve exceptionally fast practical performance while preserving good theoretical bounds for graph applications. 
In the following, we review the interface of the \ourtree in \cref{sec:interface}, followed by its algorithms in \cref{sec:operations} and analysis in \cref{sec:analysis}. 
We then present implementation details in \cref{sec:impl}, and finally compare our design with existing work in \cref{sec:discussion}. 

\subsection{Interface}\label{sec:interface}
\myparagraph{The \DataTypeA Interface.}
Here we consider the standard set interface, focusing primarily on operations often used in graph applications, listed below.
$S$ denotes the current set.
\begin{itemize}
    \item $\finit(n,b)$: initialize an empty ($b=0$) or a full ($b=1$) set over a universe of size $n$.
    \item $\fempty()$: return whether $S = \emptyset$.
    \item $\fcontains(\ind)$: return whether $\ind \in S$.
    \item $\finsert(\ind)$: (atomically) insert $\ind$ into $S$ if $\ind\notin S$; return true when successful.
    \item $\fdelete(\ind)$: (atomically) delete $\ind$ from $S$ if $\ind\in S$; return true when successful.
    \item $\fforeach(f(\cdot))$: apply function $f(\ind)$ to each $\ind\in S$. 
    \item $\freduce(f(\cdot), \oplus, \bot)$: compute $\oplus_{\ind \in S} f(\ind)$ in parallel with associative operator $\oplus$. 
    \item $\fclear()$: delete all elements in the set.
\end{itemize}

These functions capture the common needs of many parallel graph algorithms. 
To help understand the interface, we show an example of parallel BFS in \cref{algo:bfs-simple}. 
The algorithm maintains three sets: all active (unprocessed) vertices $\fa$, the frontier $\ff$, and the next frontier $\ff'$.
Each round of the algorithm applies $\fforeach$ on $\ff$ to visit neighbors ($u$) of each $v\in \ff$. 
If $\fa.\fdelete(u)$ succeeds (i.e., $u$ is discovered for the first time), we insert $u$ into $\ff'$.

\input{figures/toggle.tex}

\input{algos/pseudocode_bfs_simple.tex}

\medskip
\myparagraph{The \DataTypeB Interface.}
A natural extension of the set interface is the \emph{map} interface where each element is associated with a value. 
This is most common in distance-related graph applications, where each vertex maintains a tentative distance, 
and the algorithm often needs to extract all vertices with the smallest value (distance). 
Consequently, \DataTypeB{} also incorporates priority-queue operations. 
Without loss of generality, we assume a min-priority queue throughout the paper based on a total ordering with the following interface. 

\begin{itemize}
    \item $\finit(n)$: initialize a map over a universe of size $n$, all elements are inactive and are mapped to $+\infty$.
    \item $\fdecval(\ind,\val)$: atomically update the value of element $\ind$ to $\val$ if $\val$ is smaller, and activate it if successful.
    \item $\fextract(f(\cdot))$: let $\minval$ be the minimum value among active elements, apply $f(\ind)$ to all active elements attaining $\minval$, deactivate them and reset their values to $+\infty$, and return $\minval$.
\end{itemize}

\myparagraph{Parallelism and Concurrency.} \label{sec:concurrency}
We consider the batch-concurrent setting: For each \DataStructure instance, 
all batch operations (e.g., $\fforeach$, $\fextract$) are internally parallel and must run in isolation;
all point updates (e.g., $\finsert$, $\fdelete$, $\fdecval$) are atomic, and can only be concurrent with the same type of operations;
$\fcontains$ may run concurrently with $\finsert$ or $\fdelete$ and observe either endpoint of the corresponding membership transition: false-to-true for $\finsert$ and true-to-false for $\fdelete$.
This is common for many parallel graph algorithms, which alternate between modification (e.g., $\finsert$, $\fdelete$) and traversal (e.g., $\fforeach$, $\freduce$), 
following the frontier-based paradigm described in \cref{sec:prelim}.

\subsection{Algorithms}\label{sec:operations}

\subsubsection{The \ourtree{} Layout}

The \ourtree data structure consists of $L+1$ levels of bit vectors, 
where $L = \lceil \log_w n \rceil$. The bit vector array at the bottom level has the length of $\lceil n/w\rceil$. 
We refer to each bit as a \emph{node} in the tree. 
For ease of algorithm description, we use $\bits[i][j]$ to represent the $j$-th bit at the $i$-th level.
At the lowest level $L$, each $\bits[L][j]$ directly denotes the existence of index $j$ in the set.
For the internal nodes $\bits[i][j]$ where $i<L$, it represents the aggregation of $w$ bits at the next level (its children), namely $\bits[i+1][j\cdot w]$ to $\bits[i+1][(j+1)\cdot w-1]$. 
If all of them are 0, $\bits[i][j]=0$, otherwise $\bits[i][j]=1$. 
In the code, \emph{every group of $w$ bits is treated together as an integer}, and the efficiency of this layout comes from built-in bit operations.
In fact, checking the existence, modifying, or extracting a certain child of a node can be performed by $O(1)$
bitwise operations, instead of $O(w)$. 
We list the built-in bitwise functions we use, together with other implementation details, in \cref{sec:impl}.
In the following, we describe the algorithms in detail and analyze their costs.

\input{algos/pseudocode_IndexSet_ArrayVersion.tex}

\subsubsection{\DataTypeA{} Algorithms}

We now introduce algorithms for \DataTypeA{} and present them in \cref{algo:pvs_ops}. 

\myparagraph{$\fempty$ and $\fcontains{}(\ind)$.} By definition, $\fempty()$ simply checks the bit at the root level, while $\fcontains(\ind)$ simply checks the corresponding bit at the leaf level.

\myparagraph{$\finsert{}(\ind)$ and $\fdelete{}(\ind)$.} Both operations start from the leaf $\bits[L][\ind]$ and use $\setatomic$ to update the corresponding bit.
The implementation of $\setatomic$ is given later in \cref{algo:atomic-set} (\cref{sec:impl}).
The returned $\vname{changed}$ flag determines whether the requested set operation succeeds. 
The $\propagate$ flag is true only when the containing $w$-bit word crosses the all-zero/nonzero boundary.
In that case, the aggregate represented by the parent bit changes, so the operation continues at its parent $\bits[i-1][j/w]$.
Otherwise, upward propagation terminates early.

\myparagraph{$\fforeach(f)$.} This operation enumerates all elements in the set and applies a user-provided function $f$ to each of them. 
It calls a helper function $\fforeach^*(i,j,f)$ to process the subtree at $\bits[i][j]$, starting from the root $\bits[0][0]$. 
If $\bits[i][j]$ is a leaf ($i=L$), then it represents one active element $j$ and the algorithm executes $f(j)$. 
Otherwise, we use a parallel-for to visit all existing children of the current node, i.e., all the 1-bits of its children in the next level, 
and recursively apply $\fforeach^*$ to them. 
Note that all its children by definition are stored in a word with width $w$, and we use bitwise operations such as \emph{popcount} to enumerate all 1-bits. 

\myparagraph{$\freduce{}(\phi, \oplus,\bot)$.} This operation aggregates information over all elements in the set, with
an identity value $\bot$, a mapping function $\phi$, and an associative operator $\oplus$. 
Similar to $\fforeach$, it uses a helper function $\freduce^*(i,j,\dots)$ applied to the subtree at $\bits[i][j]$.
For a leaf node, the reduction result is simply $\phi(j)$, in which $j$ is the index of the element. 
For an internal node, the algorithm first gathers the list of children and allocates a partial array indexed by these children. 
It then recursively computes the reduction of every child in parallel, and finally combines them using $\oplus$. 
This abstraction is general enough to facilitate many set-level queries, such as counting elements, computing sums, finding minimum/maximum values, or combining user-defined summaries.

\myparagraph{$\fclear{}()$.} This function removes all elements from the set by using $\fforeach$ with $f$ deleting the element itself.

Taken together, these operations define a minimal yet expressive interface for an \DataTypeA{}. 
In \cref{sec:applications}, we will see that they efficiently support a wide range of graph algorithms.

\subsubsection{\DataTypeB{} Algorithms}
An \DataTypeB{} associates a value with each index in an \DataTypeA{}, and supports a priority-queue interface. 
The implementation in the \ourtree{} maintains three fields for each tree node: a $\bitflag$ bit, a $\dirty$ bit, and a value $\augval$.
For a leaf, $\augval$ is simply the current value of the corresponding element. 
For an internal node, $\augval$ is the minimum $\augval$ among its active children, or $+\infty$ if no child is active.
At a high level, it is similar to a parallel tournament tree with larger fan-outs. 
We present the algorithms in \cref{algo:tournament_tree}.

\myparagraph{$\fdecval{}(\ind,\val)$.} This updates the value of element $\ind$ to $\val$. It first applies an atomic $\writemin$ to the target leaf value. 
If the value decreases, ancestors on the leaf-to-root path are marked dirty by $\setatomic$ with early termination, 
similar to the process of $\finsert$ in \DataTypeA. 
Importantly, the update does not immediately recompute all subtree minima. Instead, it records that the ancestors may need a repair. 
This lazy strategy keeps concurrent updates simple and cheap. 

\myparagraph{$\fextract{}(f)$.} This applies function $f$ to all elements with the minimum value, and removes them from the set.  
The operation first invokes $\frepair^*(0,0)$ to recursively traverse dirty branches and restore the correct subtree minima. 
Once repaired, the root value is the current global minimum $\minval$, which will be the return value. 
The main procedure then calls $\fextract^*(i,j, \minval{}, f)$ with $\minval$. 
The recursion descends only into children with value no more than $\minval$. 
When it reaches a leaf, it applies $f$ to it, marks the leaf inactive and resets its value to $+\infty$. 
After the recursions, the node recomputes its value from its remaining children.

\input{algos/pseudocode_IndexMap.tex}

\myparagraph{Correctness.}
The \ourtree{} maintains the invariant that an internal bit is $1$ if and only if its child word is nonzero.
Since \faor{} and \faand{} return the previous value of the word, among concurrent updates of the same type, exactly one operation observes each zero/nonzero transition of a word and propagates it, so no update is lost and no propagation is redundant.
By induction from the leaves to the root, the invariant holds after each batch; ancestors may temporarily lag behind the leaves within a batch, which is safe because traversals are isolated from updates (\cref{sec:interface}).
For \DataTypeB{}, concurrent $\fdecval$ operations use $\writemin$ at the leaves and propagate dirty flags in the same way, and the subsequent isolated $\fextract$ repairs all dirty subtrees before extraction.

\subsection{Theoretical Analysis}\label{sec:analysis}

Intuitively, maintaining the $\DataTypeA$ using a tree instead of an array may incur a logarithmic overhead due to its tree height. 
While this applies to trees with small fan-outs, our analysis demonstrates that by setting the fan-out $w$ to the word size $\Theta(\log n)$, this overhead becomes small, or negligible in most cases.  
As a result, the $\ourtree$ can consistently deliver fast performance in graph applications due to structural simplicity, as discussed in more detail in \cref{sec:discussion}. 
Our analysis is rather simple and based on the following fact.

\begin{fact}\label{fact:tree-node}
Given a complete $w$-ary tree with $n$ leaves, the total number of nodes of $k$ leaves and their ancestors is $O\left(k\log_w\frac{n}{k}\right)$.
\end{fact}
This fact is commonly used in analyzing tree algorithms.
We can pessimistically assume that the top $\lceil \log_w k\rceil$ levels are all selected with $O(k)$ nodes.
At most $k$ nodes can be selected for the rest of $\lceil \log_w n\rceil-\lceil \log_w k\rceil=O(\log_w (n/k))$ levels, leading to the stated bound.

We can now show the bounds for \ourtree operations. 
Since some algorithms are parameterized with a function $f(\cdot)$ ($\fforeach$, $\freduce{}$, or $\fclear{}$), the actual complexity relies on the cost of $f(\cdot)$. 
To provide a general analysis, we define the \defn{accessing work/span} as the cost to traverse relevant nodes in the tree excluding the cost of these functions. 

\begin{theorem}[$\DataTypeA$ Costs]
\label{thm:indexset}
Given a \ourtree of range $n$ and fan-out~$w$, $k$ concurrent $\finsert{}$ or $\fdelete{}$ operations take $O\left(k\log_w\frac{n}{k}\right)$ work, each with $O(\log_w n)$ span. 
$\fforeach$, $\freduce{}$, or $\fclear{}$ has $O\left(k\log_w\frac{n}{k}\right)$ accessing work and $O(\log n)$ accessing span when $k$ elements are accessed.
\end{theorem}

\begin{proof}
  In a batch of same-type updates, $\setatomic$ (\cref{algo:atomic-set}) returns a true $\propagate$ flag to at most one operation on each word, so at most one operation per word continues to the parent.
  According to \cref{fact:tree-node}, since these operations require access to $k$ leaves, they will visit or modify $O\left(k\log_w\frac{n}{k}\right)$ tree nodes.
  For $\finsert{}$, $\fdelete{}$, and $\fclear{}$, all modifications to the tree nodes are performed using $\fetchandor$ and $\fetchandand$, which take constant work per operation and yield the stated work bound in the theorem.
  $\fforeach$ and $\freduce{}$ are read-only, so the work is naturally $O\left(k\log_w\frac{n}{k}\right)$.
  For $\fforeach$, $\freduce{}$, and $\fclear{}$, the traversal has at most $\lceil \log_w n\rceil$ recursive branching steps, each with $O(\log w)$ span for the parallel for-loop.
  In total, the span for these operations is $O(\log n)$ by multiplying the two terms.
\end{proof}

The cost of $\DataTypeB$ is slightly higher because its traversal cannot rely entirely on word-level bitwise operations.
Note that the accessing work of $\fextract$ accounted for in \cref{thm:indexmap} excluded the work of repairing nodes, which can be charged against the total work of $\fdecval$.

\begin{theorem}[$\DataTypeB$ Costs]
\label{thm:indexmap}
Given a \ourtree of range $n$ and fan-out~$w$, $k$ concurrent $\fdecval$ operations take $O\left(k\log_w\frac{n}{k}\right)$ work and $O(\log_w n)$ span;
$\fextract$ has $O\left(kw\log_w\frac{n}{k}\right)$ accessing work and $O(\log n)$ accessing span when the output size is $k$.
\end{theorem}

\begin{proof}
  The analysis of the $\DataTypeB$ operations is almost identical to that of $\DataTypeA$, with only one minor difference.
  In $\fextract$, since we need to check all $w$ subtrees in each relevant tree node (line~\ref{line:for-w} in \cref{algo:tournament_tree}), the work bound is increased by a factor of $w$, making it slightly less efficient than other \ourtree{} operations.
\end{proof}

The bound indicates that all operations on the \ourtree{} are roughly $O\left(\log_w\frac{n}{k}\right)$ per element, 
which can be $O(\log n / \log \log n)$ (i.e., the tree depth) when plugging in $k=O(1)$ and $w=\Theta(\log n)$.
We note that, however, for tasks suitable for parallel graph processing, the set sizes are usually reasonably large to enable meaningful parallelism. 
Motivated by this observation, we provide a more detailed analysis of the cost of using \ourtree{s} for graph algorithms, which can be directly plugged in the analysis for the applications using the \ourtree in \cref{sec:applications}.

\begin{theorem}\label{thm:rounds}
  To process $s_{\mathit{sum}}$ elements in a \ourtree in a total of $\rho$ rounds, the accessing work is $O\left(s_{\mathit{sum}}\log_{w}\frac{n\rho}{s_{\mathit{sum}}}\right)$ work, or $O\left(s_{\mathit{sum}}\cdot\frac{\log({n\rho}/{s_{\mathit{sum}}})}{\log\log n}\right)$ when $w=\Theta(\log n)$.
\end{theorem}

\begin{proof}
  Let $s_1, s_2, \ldots, s_\rho$ be the number of elements accessed in each of the $\rho$ rounds, where $\sum_{i=1}^\rho s_i = s_{\mathit{sum}}$. 
  The accessing work is bounded by $O\left(\sum_{i=1}^\rho{s_i\log_w\frac{n}{s_i}}\right)$. 
  Because the function $f(x) = x\log_w(n/x)$ is concave for $x > 0$, by Jensen's inequality, this total cost is maximized in the worst case when the elements are evenly distributed, i.e., $s_1=s_2=\cdots=s_\rho=s_{\mathit{sum}}/\rho$. 
  Plugging in these values yields the stated bound.
\end{proof}

With \cref{thm:rounds}, we can derive tighter bounds for graph algorithms.
For instance, in breadth-first search (BFS), every vertex is visited at most once, which means $s_{\mathit{sum}}\le n$.
Since the bound in \cref{thm:rounds} is non-decreasing in $s_{\mathit{sum}}$, plugging it in yields $O\left(n \cdot \frac{\log \rho}{\log\log n}\right)$ accessing work, where $\rho$ is the number of BFS rounds. 
For graph processing tasks that may benefit from parallelism, the number of rounds is usually small. 
For example, if $\rho$ is polylogarithmic, the work becomes $O\left(n \cdot \frac{\log \polylog(n)}{\log\log n}\right)=O(n)$, which is optimal.

\subsection{Implementation Details}\label{sec:impl}

\myparagraph{Layout and Initialization.}
When $n$ is not a power of $w$, the implementation does not materialize a complete tree.
Instead, only the last word of each level may contain padding.
These padding bits are set to $0$ so that they never represent indices or nonempty subtrees.
\Cref{algo:indexset-init} gives the corresponding initialization procedure.

\begin{algorithm}[!htbp]
\small
    \caption{Implementation of $\DataTypeA.\finit(n,b)$\label{algo:indexset-init}}
    \DontPrintSemicolon
    \SetKwProg{myfunc}{Function}{}{}
    \SetKw{KwDownTo}{downto}
    \myfunc{\upshape$\finit(n,b)$}{
        $L \gets \lceil log_w n\rceil$,~$q \gets n$ \tcp*[f]{number of valid bits}\\
        \For{$i\gets L$ \KwDownTo $0$}{
            $x \gets 2^w-1$ if $b=1$, and $0$ otherwise\\
            $\bits[i] \gets$ an array of $\lceil q/w\rceil$ words initialized to $x$\\
            \tcp{clear padding bits}
            \If{$b=1 \land (q \bmod w)\ne 0$}{
                $\bits[i][\lceil q/w\rceil-1] \gets (1 \ll (q \bmod w))-1$ 
            }
            $q \gets \lceil q/w\rceil$ \tcp*[f]{valid parent bits}
        }
        \Return $\bits$
    }
\end{algorithm}

\myparagraph{Bitwise Operations for Updates.}
\Cref{algo:atomic-set} gives the $\setatomic$ primitive used for point updates, where literal $1$ denotes the $w$-bit unsigned integer $1$.

\begin{algorithm}[!htbp]
\small
    \caption{Implementation of $\setatomic$\label{algo:atomic-set}}
    \DontPrintSemicolon
    \SetKwProg{myfunc}{Function}{}{}
    \small
    \tcp{atomically set $\bits[i][j]$ to $b$ and return $(\vname{changed}, \propagate)$}
    \myfunc{\upshape$\setatomic(\bits[i][j],b)$}{
        $W \gets$ the $w$-bit word containing $\bits[i][j]$ \\
        $m \gets 1 \ll (j \bmod w)$ \tcp*[f]{mask of the target bit}\\
        \eIf{$b=1$}{
            $old \gets \faor(W,m)$ \\
            \tcp{whether $\bits[i][j]$ was originally 0}
            $\vname{changed} \gets ((old \mathbin{\&} m)=0)$ \\
            \tcp{whether $W$ toggled from all-zero to nonzero}
            $\propagate \gets (old=0)$
        }{
            $old \gets \faand(W,\mathord{\sim}m)$ \\
            \tcp{whether $\bits[i][j]$ was originally 1}
            $\vname{changed} \gets ((old \mathbin{\&} m)\ne 0)$ \\
            \tcp{whether $W$ toggled from nonzero to all-zero}
            $\propagate \gets (old=m)$
        }
        \Return $(\vname{changed}, \propagate)$
    }
\end{algorithm}

\myparagraph{Bitwise Operations for Traversal.}
To ensure that word-level operations require only $O(1)$ work, we employ three GCC built-ins:

\begin{itemize}
  \item \popcount{} ({\insformat{\_\_builtin\_popcountll}}) returns the number of 1-bits in a 64-bit word.
  \item \ctz{} ({\insformat{\_\_builtin\_ctzll}}) returns the index of the lowest 1-bit in a nonzero 64-bit word.
  \item \pdep{} ({\insformat{\_\_builtin\_ia32\_pdep\_di}}, from the x86-64 BMI2 instruction set) satisfies the following property, the only one we use: if the $i$-th $1$-bit of a word is at bit position $x$, then
$\texttt{\pdep(1}<\!\!<\texttt{i,word) = 1}<\!\!<\texttt{x}$.
\end{itemize}

Thus, for a given word, \popcount{} determines the iteration range of a parallel loop,
and the $i$-th task obtains the offset of its set bit in $O(1)$ work as \texttt{\ctz(\pdep(1}$<\!\!<$\texttt{i,word))}.

\myparagraph{Granularity Control.}
To amortize parallel overhead, we apply granularity control only at the last level: among the $w=64$ elements of a word, we group every eight \emph{active} elements into a single task, rather than assigning a dedicated task to each element.
In our implementation, this simple strategy performs well on almost all graphs.

\subsection{Comparisons to Existing Approaches}
\label{sec:discussion}

In this section, we briefly review typical frontier representations in existing graph algorithms, especially the baselines in our experiments, 
and discuss how \ourtree{} is designed to address the challenges in them. We still use the example of BFS algorithm in \cref{algo:bfs-simple}. 
As mentioned, the most widely-used form is to materialize all vertices in a contiguous array, as is used in many graph libraries~\cite{shun2013ligra,gbbs2021,dhulipala2020semi,dhulipala2017}. 
This streamlines $\fforeach$ traversals, but prevents concurrent insertions to the next frontier, because appending to a list of indices is not thread-safe, and a shared atomic append counter incurs heavy contention. 
This can be addressed by a two-pass traversal of $\ff$: first to mark the vertices to be inserted, and the second time to pack them into $\ff'$, which is a standard race-free solution~\cite{shun2013ligra,gbbs2021}. 
Another solution, represented by the \emph{\hashbag}~\cite{wang2023parallel}, uses a hash-table-like strategy: it leaves sufficient empty slots, and a concurrent insert puts an element to a random slot in the array. 
Both of them incur undesired overhead in time and/or space in practice. 

Beyond the sparse representation, almost all existing parallel BFS implementations incorporate a dense mode, where the presence of a vertex in the set is indicated by a bit flag. 
Extensive experimental studies~\cite{Beamer12,shun2013ligra,zhang2018graphit,dong2021efficient} have demonstrated the effectiveness of the dense mode in accelerating parallel graph processing.
The advantage stems in part from its simplicity, and additionally from its facilitation of the \emph{direction optimization}~\cite{Beamer12} (see \cref{algo:bfs} in \cref{sec:applications}). 
However, the $\fforeach$ operations in the dense mode require checking all vertices and processing the flagged ones. 
This restricts the applicability of the dense representation only to a few rounds with large frontiers. 

The representation of \ourtree{} is also based on bit flags, making its update cost close to the dense mode. 
Given $w = 64$ on modern machines, the tree height is at most 7 for any graph that fits in memory (up to $64^6 \approx 6.9 \times 10^{10}$ vertices). 
Therefore each single update modifies at most 7 bits, and can be much fewer with early termination (see \cref{sec:analysis}). 
For traversal, it does not exhaustively check all bits, but uses the hierarchy to skip subtrees with all 0s.
This is the motivation for the hierarchy: a flat bitmap must scan all $n$ bits even for a small set, whereas the upper levels of a \ourtree{} are small, frequently accessed and thus likely in cache, adding little overhead (see \cref{sec:exp:simple}). 
This makes the traversal performance competitive with the sparse mode, with a minor overhead factor of $O(\log_w n/k)$ for subset size $k$. 
The value $\log_w n/k$ is also bounded by 7, and can be smaller for larger $k$.
Altogether, based on \cref{thm:rounds}, when a graph algorithm finishes in $\polylog(n)$ rounds, the overhead is a constant factor.

Using \ourtree{}, an algorithm does not need to distinguish sparse and dense modes. 
Beyond BFS, for problems such as graph coloring and \kcore{}, many state-of-the-art implementations never use a dense mode, 
because the frontiers rarely grow sufficiently large. Using \ourtree{}, they can also leverage fast updates as if in dense modes, 
without much penalty in the theoretical efficiency to traverse the frontiers. 
In our experiments, our code for these two algorithms using the \ourtree{} indeed improves existing solutions.

%% file: figures/toggle.tex
\begin{figure}[t]
\begin{center}
\includegraphics[width=\columnwidth]{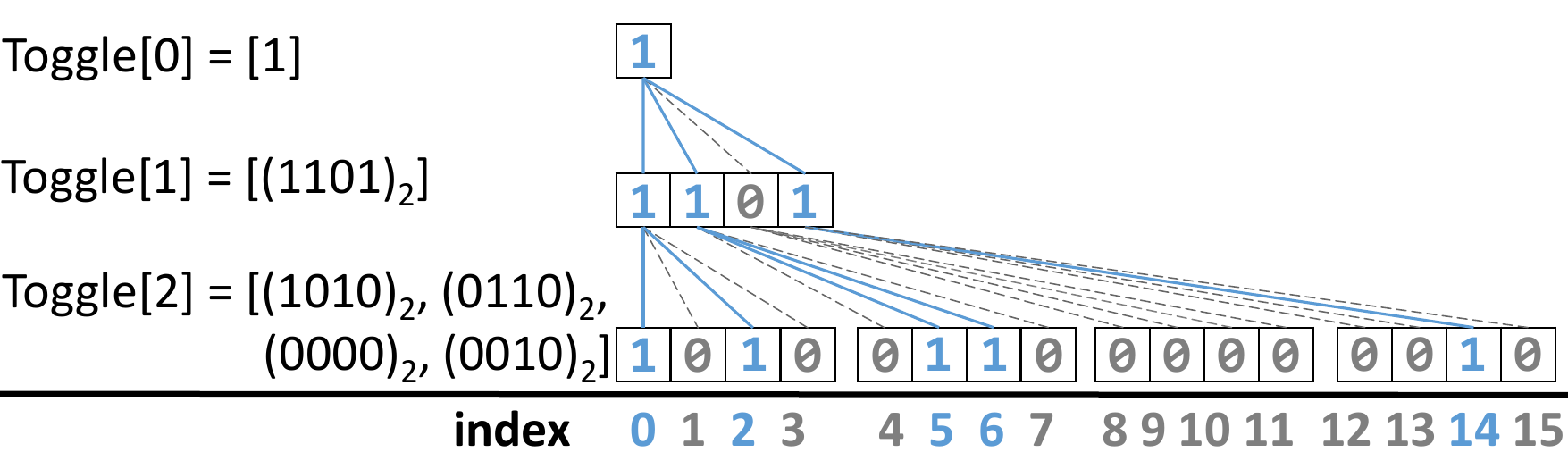}
\captionof{figure}{\textbf{\boldmath An illustration of an \DataTypeA{} $\{0,2,5,6,14\}$ represented by a \ourtree{}.\unboldmath} Here we use $n=16$ and $w=4$. }
\label{fig:toggle}
\end{center}
\end{figure}

%% file: algos/pseudocode_bfs_simple.tex
\begin{algorithm}[t]
\small
    \caption{Parallel BFS using the \DataTypeA{} interface \label{algo:bfs-simple}}
    \SetKwFor{EmptyLoop}{}{do}{}
    \SetKwFor{parfor}{ParallelForEach}{do}{}
    \DontPrintSemicolon
    \myfunc{\upshape$\textsf{BFS}(G=(V,E),\ \vname{source}\in V)$}{
        \tcp{the active (unprocessed) vertices}
        \tcp{$\fa$ is never enumerated here; a flat bitmap also suffices}
        $\fa \gets \DataTypeA.\finit(G.n,\ 1)$, $\fa.\fdelete(source)$ \\
        $\ff \gets \DataTypeA.\finit(G.n,\ 0)$, $\ff.\finsert(source)$ \tcp*[f]{frontier}\\   
        $\ff' \gets \DataTypeA.\finit(G.n,\ 0)$\tcp*[f]{initialize the next frontier}\\
        \While{\upshape$\neg~\ff.\fempty()$}{
            \hspace{-1mm}
            \EmptyLoop{\upshape$\ff.\fforeach \ v \in \ff$}{
                \parfor{$u \in G.N(v)$}{
                    \If(\tcp*[f]{first discovery of $u$}){\upshape$\fa.\fdelete(u)$}{
                        $\ff'.\finsert(u)$ \tcp*[f]{add to the next frontier}\\
                    }
                }
            }
            $\ff.\fclear{}()$\\
            swap $\ff$ and $\ff'$\\
        }
    }
    
\end{algorithm}

%% file: algos/pseudocode_IndexSet_ArrayVersion.tex
\begin{algorithm*}[htbp]
\small
\fontsize{9pt}{10.25pt}\selectfont 
    \caption{Algorithms for \DataTypeA \label{algo:pvs_ops}}
    \DontPrintSemicolon
    \SetKwProg{myfunc}{Function}{}{}
    \SetKwFor{EmptyLoop}{}{do}{}
    \SetKwFor{parfor}{ParallelForEach}{do}{}
    \SetKw{KwDownTo}{downto}

    \SetKwInput{datastructure}{{Data Structure}}
    \SetKwInput{notes}{{Notes}}
    \SetKwRepeat{Repeat}{repeat}{until}
    
    \datastructure{\\
    \begin{adjustwidth}{0.3cm}{0cm}
    \noindent $L+1$ levels of bit vectors, where $L=\lceil\log_w n \rceil$. 
    For ease of algorithm description, we use $\bits[i][j]$ to represent the $j$-th bit at level $i$, where $0\le i\le L$, and $0\le j < w^i$. 
    In the actual implementation, every $w$ bits are stored as an integer. 
    \end{adjustwidth}
    }
    \notes{\\
    \begin{adjustwidth}{0.3cm}{0cm}
    \noindent 
    \setatomic{}$(\bits[i][j],b)$ means to atomically set the bit at $\bits[i][j]$ to  $b \in \{0,1\}$ and returns a boolean pair $(\vname{changed}, \propagate)$.
    Here, $\vname{changed}$ indicates whether the target bit changes, while $\propagate$ indicates whether the containing $w$-bit word crosses the all-zero/nonzero boundary:
    from all-zero to nonzero when $b=1$, or from nonzero to all-zero when $b=0$.
    It is implemented using hardware-supported atomic bitwise operations \faor{} / \faand{} on the corresponding word with a proper bitmask. 
    \end{adjustwidth}
    }
    \begin{multicols}{2}
    \raggedcolumns
   
    \myfunc(\tcp*[f]{checks whether the root is 0}){\upshape $\fempty()$}{\Return $\bits[0][0] = 0$}
    \codeskip
    
    \myfunc(\tcp*[f]{checks the corresponding bit in the leaf}){\upshape$\fcontains(\ind)$}{\Return $\bits[L][\ind] = 1$}
    \smallskip
    \myfunc(){\upshape $\finsert(\ind)$}{
        $i\gets L$, $j\gets \ind$ \tcp*[f]{start from the index at the leaf level}\\
        $(\vname{changed}, \propagate) \gets \setatomic(\bits[i][j],1)$\\
        \lIf{$\neg$~$\vname{changed}$}{\Return{\false{}}}
        \While{$i>0 \land \propagate$}{
            $i\gets i-1$, $j\gets j/w$ \tcp*[f]{get the parent of $\bits[i][j]$}\\
            $(\_, \propagate) \gets \setatomic(\bits[i][j],1)$\\
        }
        \Return{\true{}}
    }
    \codeskip
    \myfunc{\upshape$\fdelete(\ind)$}{
        $i\gets L$, $j\gets \ind$ \tcp*[f]{start from the index at the leaf level}\\
        $(\vname{changed}, \propagate) \gets \setatomic(\bits[i][j],0)$\\
        \lIf{$\neg$~$\vname{changed}$}{\Return{\false{}}}
        \While{$i>0 \land \propagate$}{
            $i\gets i-1$, $j\gets j/w$ \tcp*[f]{get the parent of $\bits[i][j]$}\\
            $(\_, \propagate) \gets \setatomic(\bits[i][j],0)$\\
        }
        \Return{\true{}}
    }
    
    \codeskip
    \myfunc(\tcp*[f]{remove all indices}){\upshape$\fclear()$}{
        $\fforeach^{*}(0,0,\ind\mapsto \fdelete(\ind))$
    }

    \codeskip
        
    \myfunc(\tcp*[f]{apply $\vname{f}(\ind)$ to all indices in the set}){\upshape$\fforeach(\vname{f})$}{
      $\fforeach^{*}(0,0,\vname{f})$\tcp*[f]{Use helper function $\fforeach^{*}$}
    }
    \codeskip
    \tcp{apply function $\vname{f}(\ind)$ to $\bits[i][j]$ and all its descendants}
    \myfunc{\upshape$\fforeach^*(i,j,\vname{f})$}{
      \lIf{$i=L$}{$\vname{f}(j)$}
      \Else(\tcp*[f]{recursively process all existing children of $\bits[i][j]$}){
        $C \gets \{j^*: w \cdot j \le j^* < w \cdot (j+1), \bits[i+1][j^*] = 1\}$ \\
        \parfor{$j^* \in C$}{
          {$\fforeach^{*}(i+1,j^*,\vname{f})$}
        }
      }
    }
    \codeskip
    \tcp{for all indices in the set, reduce $\vname{\phi}(\ind)$ using an associative operator $\vname{\oplus}$ and identity $\bot$}
    \myfunc{\upshape$\freduce(\vname{\phi}, \vname{\oplus},\bot)$}{
      \Return $\freduce^{*}(0,0,\vname{\phi}, \vname{\oplus},\bot)$\tcp*[f]{Use helper function $\freduce^{*}$}
    }
    \codeskip
    \tcp{apply $\freduce(\vname{\phi}, \vname{\oplus},\bot)$ to the subtree at $\bits[i][j]$}
    \myfunc{\upshape$\freduce^{*}(i,j, \vname{\phi},\vname{\oplus},\bot)$}{
        \lIf{$i=L$}{
            \Return $\vname{\phi}(j)$
        }
        \Else(\tcp*[f]{recursively process all existing children of $\bits[i][j]$}){
            $C \gets \{j^*: w \cdot j \le j^* < w \cdot (j+1), \bits[i+1][j^*] = 1\}$ \\
            \lIf {$C = \emptyset$}{\Return{$\bot$}}
            \parfor{$j^* \in C$}{
                {$t[j^*-w\cdot j]\gets \freduce^{*}(i+1,j^*, \vname{\phi},\vname{\oplus},\bot)$}
            }
            \Return{$\oplus_{j^* \in C}t[j^*-w\cdot j]$}
        } 
    }

    \end{multicols}
\end{algorithm*}

%% file: algos/pseudocode_IndexMap.tex
\begin{algorithm}[!t]
\fontsize{9pt}{10.25pt}\selectfont 
    \caption{Algorithms for \DataTypeB\label{algo:tournament_tree}}
    \DontPrintSemicolon
    \SetKwProg{myfunc}{Function}{}{}
    \SetKwFor{EmptyLoop}{}{do}{}
    \SetKwFor{parfor}{ParallelForEach}{do}{}
    \SetKwInput{datastructure}{{Data Structure}}
    
    \datastructure{\\
    \begin{adjustwidth}{0.3cm}{0cm}
    \noindent $(L+1)$ levels of toggles, where $L=\lceil\log_w n \rceil$. 
    For ease of algorithm description, we use $\bits[i][j]$ to represent the $j$-th node at level $i$, where $0\le i\le L$, and $0\le j < w^i$. 
    Each toggle has a $\bitflag$ bit to indicate its existence during $\fextract^*$, a $\dirty$ bit to indicate the staleness of the value, and a $\augval$ as its value. 
    \end{adjustwidth}
    }
    \smallskip
    \tcp{decrease the value of $\ind$ and mark ancestors dirty}
    \myfunc{\upshape$\fdecval(\ind, \val)$}{
        \If{\upshape$\writemin(\bits[L][\ind].\augval, \val)$}{
            $i\gets L$, $j\gets \ind$ \\
            $(\_, \propagate) \gets \setatomic(\bits[i][j].\dirty, 1)$\\
            \While{$i>0 \land \propagate$}{
                $i\gets i-1$, $j\gets j/w$ \\
                $(\_, \propagate) \gets \setatomic(\bits[i][j].\dirty, 1)$\\
            }
        }
    }
    \smallskip
    \tcp{apply $f$ to all minimum-valued elements and remove them}
    \myfunc{\upshape$\fextract(\vname{f})$}{
        \lIf{$\bits[0][0].\dirty = 1$}{
            $\frepair^*(0,0)$
        }
        $\minval \gets \bits[0][0].\augval$ \\
        \lIf{$\bits[0][0].\bitflag = 1$}{
            $\fextract^*(0,0,\minval,\vname{f})$
        }
        \Return $\minval$
    }

    \tcp{restore the minimum value in the dirty subtree $\bits[i][j]$}
    \myfunc{\upshape$\frepair^*(i,j)$}{
        $\bits[i][j].\dirty \gets 0$, $\bits[i][j].\bitflag \gets 1$ \\
        \lIf{$i=L$}{
            \Return 
        }
        $C \gets \{j^*: w \cdot j \le j^* < w \cdot (j+1), \bits[i+1][j^*].\dirty=1\}$ \\
        \parfor{$j^* \in C$}{
            $\frepair^*(i+1,j^*)$\\
            write\_min$(\bits[i][j].\augval,\bits[i+1][j^*].\augval)$ \\
        }
    }
    \medskip
    \tcp{find all active elements in subtree $\bits[i][j]$ with minimum value $\minval$, apply $f$ to them, and remove them}
    \myfunc{\upshape$\fextract^*(i,j, \minval, \vname{f})$}{
        \If(\tcp*[f]{reaching the leaf level}){$i=L$}{
            $\vname{f}(j)$ \tcp*[f]{apply $f$ to it}\\
            $\bits[i][j].\bitflag \gets 0$,~$\bits[i][j].\augval \gets +\infty$\\
            \Return
        }
        \tcp{enumerate all children with $\minval$}
        $C \gets \{j^*: w \cdot j \le j^* < w \cdot (j+1), \bits[i+1][j^*].\bitflag = 1\}$ \\
        \parfor{
            $j^* \in C, \bits[i+1][j^*].\augval = \minval$\label{line:for-w}
        }{
            $\fextract^*(i+1,j^*,\minval,\vname{f})$
        }
        \tcp{recalculate the value as the minimum value in its subtree}
        $C \gets \{j^*: w \cdot j \le j^* < w \cdot (j+1), \bits[i+1][j^*].\bitflag = 1\}$ \\
        \lIf {$C = \emptyset$}{$\bits[i][j].\bitflag \gets 0$,~$\bits[i][j].\augval \gets +\infty$}
        \lElse {$\bits[i][j].\augval \gets \min_{j^* \in C}\bits[i+1][j^*].\augval$}
    }
\end{algorithm}

%% file: main/04_application.tex
\section{Graph Applications using the \ourtree{}}\label{sec:applications}

In this section, we discuss how to apply \ourtree{} to five well-known parallel graph algorithms to demonstrate its usability.
Integrating \ourtree{} into these algorithms is straightforward: we keep the algorithms unchanged and maintain vertex subsets using \ourtree{s}. 
As we will show in \cref{sec:exp}, \ourtree{} significantly improves the performance of these algorithms compared to existing baselines.

\input{tables/algo_workspan.tex}
\input{algos/pseudocode_bellmanford.tex}
\medskip
\myparagraph{Bellman-Ford.}
The Bellman-Ford algorithm computes single-source shortest paths from a $\source$ vertex. 
The algorithm maintains a frontier (subset of vertices) in each round and updates their neighbors.
If the distance of a vertex is updated, it is added to the next frontier.
The pseudocode is shown in \cref{algo:bellmanford}.
We use two alternating \ourtree{s} to represent the current and next frontiers, denoted as $\ff$ and $\ff'$.
The frontier size can be $n$ in the worst case, leading to the following bounds (as good as any known implementation).

\begin{corollary}
In a graph with $n = |V|$ vertices and $m = |E|$ edges, \cref{algo:bellmanford} (Bellman-Ford) has
$O\left(\rho(m+n)\right)$ work and $O(\rho\log n)$ span, where $\rho$ is the number of rounds executed.
\end{corollary}

\input{algos/pseudocode_kcore.tex}

\myparagraph{\kcore{} Decomposition.}
The $k$-core of a graph is the maximal induced subgraph in which every vertex has degree at least $k$.
A $k$-core decomposition computes the \emph{coreness} of every vertex (see \cref{fig:kcore}), where $\coreness[v]$ is the maximum $k$ such that $v$ is in the $k$-core of the graph.

The $k$-core algorithm is a typical \emph{peeling} algorithm.
In each \emph{round}, the algorithm sets $k$ to the minimum degree among the unpeeled vertices, and \emph{peels} (removes) all vertices with degree $k$ in parallel.
This may cause the degrees of remaining vertices to drop to $k$,
and the process repeats until all vertices have degree greater than $k$, forming several \emph{subrounds}.
All vertices peeled in this round have coreness of $k$.
The algorithm then continues to the next round.
To distinguish from the degree in the original graph, we refer to the modified degree in the algorithm as the \emph{induced} degree.
Here we assume the total number of subrounds is $\rho$~\cite{dhulipala2017,liu2025parallel}.

We present the $k$-core algorithm using \ourtree{s} in \cref{algo:kcore}, originally from~\cite{dhulipala2017,liu2025parallel}, with an illustration in \cref{fig:kcore}.
We maintain three vertex subsets, the active set $\fa$ (blue, all unpeeled vertices), the current frontier $\ff$ (yellow, being peeled), and the next frontier $\ff'$ (to be peeled in the next subround), each is an \DataTypeA{} implemented by a \ourtree.
$\ff$ and $\ff'$ are similar to the two alternating frontiers in Bellman-Ford. $\fa$ is initialized as full, and once a vertex is processed in the frontier, it is removed from $\fa$ (\cref{line:kcore-remove}).
In each round, $k$ is computed as the minimum induced degree among vertices in $\fa$.
Vertices with degree $k$ are identified in $\fa$ and inserted into the initial frontier of the round.
Processing a vertex $v\in\ff$ also attempts to decrement the induced degree of each neighbor that remains active.
If a neighbor $u$ drops from induced degree $k+1$ to $k$, it is inserted into the next frontier $\ff'$.
At the end of each subround, the algorithm clears $\ff$ and swaps $\ff$ with $\ff'$.

The original algorithm from~\cite{liu2025parallel} has $O(m+n)$ work and $O(\rho \log n)$ span.
This bound assumes that extracting the vertices from $\fa$ takes $O(|\fa|)$ work, which can be achieved using a flat array, with each extraction implemented by a parallel \emph{pack/filter}.
Using \ourtree{s}, however, the work is slightly higher due to the hierarchical structure. 
By using the \ourtree{}, we can achieve the following bound according to \cref{thm:rounds}, which is work-efficient when $\rho=\polylog(n)$. 

\begin{corollary}\label{cor:kcore}
Given a graph with $n$ vertices and $m$ edges, let $\rho$ be the number of subrounds, \cref{algo:kcore} ($k$-core decomposition) has $O\left((m+n) \frac{\log \rho}{\log \log n}\right)$ work and $O(\rho \log n)$ span.
\end{corollary}

\input{theorems/corollary_kcore.tex}

Although the algorithm has a multiplicative overhead of $O(\frac{\log\rho}{\log\log n})$ in work, this value is small in practice.
For $\rho=\vname{polylog(n)}$, this overhead is constant.
Benefiting from its structural simplicity, our algorithm actually outperforms Liu et al.'s flat-array implementation in the experiments.

\myparagraph{Breadth-First Search (BFS).}
We have already shown how a simple BFS uses \ourtree{s} in \cref{algo:bfs-simple} in \cref{sec:interface}.
The advanced version of parallel BFS also includes a technique called direction optimization~\cite{Beamer12}, which is used in existing implementations~\cite{shun2013ligra,gbbs2021,dong2024pasgal,beamer2015gap}. 
We present the direction-optimized BFS using \ourtree{s} in \cref{algo:bfs}.

\input{algos/pseudocode_bfs.tex}

This algorithm also keeps two alternating frontiers $\ff$ and $\ff'$, and an active set $\fa$ that maintains unvisited vertices, each implemented by a \ourtree.
The only algorithmic difference from \cref{algo:bfs-simple} is that, when a frontier is sufficiently large, a \emph{backward} strategy is used, in which each active vertex $v\in\fa$ checks its neighbors,
and if a neighbor is not in the active set $\fa$, $v$ adds itself to $\ff$. 
Here, switching between the two modes is performed by the \freduce{} function (we use fixed thresholds $\alpha=1/8$ and $\beta=1/64$ in experiments), on lines~\ref{line:bfs-switch-1} and~\ref{line:bfs-switch-2}, using the same heuristics as in the existing algorithms.

The forward mode uses a slightly different discovery pattern from \cref{algo:bfs-simple}.
The simplified algorithm attempts $\fa.\fdelete(u)$ for every examined edge and uses its return value to identify the thread that first discovers $u$.
In contrast, \cref{algo:bfs} first calls $\fa.\fcontains(u)$, avoiding atomic update attempts for vertices that are already inactive.
Under the concurrency semantics in \cref{sec:interface}, $\fa.\fcontains(u)$ on \cref{line:bfs-concurrency1} may overlap with $\fa.\fdelete(u)$.
Consequently, multiple threads may observe the same vertex $u$ as active and execute the updates on line~\ref{line:bfs-concurrency2}.
These duplicate updates are benign: after the parallel phase, $u$ is absent from $\fa$ and appears exactly once in $\ff'$, regardless of how many threads attempted the updates.
Vertices that became inactive in an earlier round are always observed as inactive and are never reinserted.
There are at most $O(m)$ duplicate update attempts, so this optimization does not change the work bound.

Based on \cref{thm:rounds}, the work and span are stated below.
Here the work is optimal when $\rho=\polylog(n)$ or $m=\Omega\left(n\frac{\log\rho}{\log\log n}\right)$.

\begin{corollary}
\label{col:bfs}
Given a graph with $n$ vertices, $m$ edges, and diameter $\rho$, \cref{algo:bfs} (BFS) has $O\left(m + n \cdot \frac{\log \rho}{\log\log n}\right)$ work and $O(\rho \log n)$ span. 
\end{corollary}

\smallskip
\begin{proof}
In the forward mode, there are at most $m$ edge visits, and the total number of vertex accesses is at most $n$, so the work is $O\left(m + n + n \cdot {\log \rho}/{\log\log n}\right)$.
In the backward mode, although a single round could visit up to $m$ edges and perform up to $n$ insertions into the frontier, the algorithm remains in this mode for at most $1/\beta$ rounds,
so it does not change the total work bound. 
\end{proof}

\input{algos/pseudocode_coloring.tex}

\myparagraph{Graph Coloring.}
The graph coloring problem requires assigning a color to every vertex such that no two adjacent vertices share the same color.
Here we consider a commonly used parallel algorithm based on the largest-degree-first heuristic~\cite{gbbs2021,hasenplaugh2014ordering}, where vertices with the largest degree are colored first, using the smallest possible color that is not used by their colored neighbors.
This iterative process finishes when all vertices are colored.
The algorithm based on the \ourtree{} is given in \cref{algo:coloring}, where we also keep two alternating frontiers $\ff$ and $\ff'$ for the current vertices being colored, and those in the next round.
The algorithm keeps $\vname{priority}[v]$ for each vertex $v$, indicating the number of uncolored neighbors that have higher degrees (breaking ties based on a random permutation).
Each vertex in the current frontier decrements its neighbors' counters, and adds them to the next frontier when the counters become 0.
\Cref{line:coloring-color1,line:coloring-color2} abbreviate the coloring procedure used in GBBS~\cite{gbbs2021}; therefore, \cref{algo:coloring} inherits its span analysis.
Based on \cref{thm:rounds}, we can derive the following work bound, which is optimal when  $\rho=\polylog(n)$ or $m=\Omega\left(n\cdot\log\rho/\log\log n\right)$.

\begin{corollary}
Given a graph with $n$ vertices and $m$ edges, \cref{algo:coloring} (degree-based coloring) has 
$O\left(m + n \cdot \frac{\log \rho}{\log\log n}\right)$ work and $O(\rho\log n)$ span, where $\rho$ is the number of rounds executed by the algorithm (or the priority DAG depth).
\end{corollary}

\input{algos/pseudocode_wBFS.tex}

\myparagraph{\titlecap{weighted-BFS.}}
Weighted-BFS (wBFS) computes single-source shortest paths on positive integer-weight graphs.
\Cref{algo:wbfs1} shows two algorithms using \DataTypeA{} and \DataTypeB{} respectively.
When using \DataTypeA{}, we keep the active set $\fa$ for discovered but unsettled vertices.
In each round, we loop over $\fa$, move those with the smallest tentative distance to $\ff$, mark them as settled, and relax their neighbors (inserting them back to $\fa$ if successful).
For the \DataTypeB{} version, we directly call $\fextract$ to find all vertices with smallest tentative distance from $\fa$.
When neighbors are successfully relaxed, we submit the modification to the priority queue using $\fdecval$.
We state the work and span bounds below, followed by their analysis.

\begin{corollary}
\label{col:wbfs}
Given a graph with $n$ vertices and $m$ edges, \cref{algo:wbfs1} (weighted-BFS) has $O\left(m + n \rho\right)$ work and $O(\rho \log n)$ span based on \DataTypeA; $O\left((m + n \log n) \frac{\log \rho}{\log\log n}\right)$ work and $O(\rho \log n)$ span based on \DataTypeB .
\end{corollary}

\begin{proof}

For the \DataTypeA{} version of \cref{algo:wbfs1}, since each vertex appears in the frontier at most once, the work for edge relaxations is $O\left(m + n + n \cdot \frac{\log \rho}{\log\log n}\right)$.
Another part of the work is extracting the vertices with minimum distance, which depends on how many vertices are active across all rounds.
This quantity cannot be bounded more tightly in general, so we use the upper bound $s_{\mathit{sum}} = n \rho$. Thus, the work bound of \cref{algo:wbfs1} is
$O\left(m + n + n \cdot \frac{\log \rho}{\log\log n} + n \rho + n \rho \cdot \frac{\log (\frac{n\rho}{n \rho})}{\log\log n}\right) = O\left(m + n \rho\right)$.

As for the \DataTypeB{} version of \cref{algo:wbfs1}, the settling property of Dijkstra's algorithm on positive-weight graphs ensures that, once a vertex $v$ is extracted, $\vname{dist}[v]$ is final and
cannot be decreased by subsequent relaxations. Hence, each vertex is extracted at most once, each edge causes at most one $\writemin$ attempt, and at most $m$ successful
relaxations invoke $\fdecval$. So the work for edge relaxations is now $O\left((m + n) \frac{\log \rho}{\log\log n}\right)$.
For extracting the vertices with minimum distance, we have $s_{\mathit{sum}} = n$, and thus the work of this part is
$O\left(n \log n (1 + \frac{\log \rho}{\log\log n})\right)$.
In all, the work bound of the \DataTypeB{} version is

\noindent $O\left(m \cdot \frac{\log \rho}{\log\log n} + n \log n (1 + \frac{\log \rho}{\log\log n})\right)$, 

\noindent with $O\left((m + n \log n) \frac{\log \rho}{\log\log n}\right)$ as shorthand.
\end{proof}

%% file: tables/algo_workspan.tex
\begin{table}[t]
  \centering
  \resizebox{\columnwidth}{!}{%
  \setlength{\tabcolsep}{0pt}
  \renewcommand{\arraystretch}{0.95}
  \begin{tabular}{@{}lccc@{}}
    \toprule
    \textbf{Algorithm} & \textbf{Rounds ($\rho$)} & \textbf{Work} & \textbf{Span} \\
    \midrule
    Bellman-Ford
      & \hspace{2pt}shortest-path tree depth
      & $O\left(\rho(m+n)\right)$
      & $O(\rho\log n)$ \\[6pt]
    BFS
      & diameter
      & $O\left(m + n \cdot \frac{\log \rho}{\log\log n}\right)$
      & $O(\rho\log n)$ \\[6pt]
    Coloring
      & priority DAG depth
      & $O\left(m + n \cdot \frac{\log \rho}{\log\log n}\right)$
      & $O(\rho\log n)$ \\[6pt]
    $k$-core
      & peeling rounds
      & $O\left((m+n) \frac{\log \rho}{\log\log n}\right)$
      & $O(\rho\log n)$ \\[6pt]
    weighted-BFS
      & weighted diameter
      & $O\left((m + n \log n) \frac{\log \rho}{\log\log n}\right)$
      & $O(\rho\log n)$ \\[3pt]
    \bottomrule
  \end{tabular}
  }
  \caption{\textbf{Work and span of graph algorithms using \DataStructure.} 
    $m$: number of edges;
    $n$: number of vertices;
    $\rho$: number of rounds. We assume the machine word width $w=\Theta(\log n)$.}
  \label{table:algo_workspan}
\end{table}

%% file: algos/pseudocode_bellmanford.tex
\begin{algorithm}[t]
\small
    \caption{Parallel Bellman-Ford using \DataStructure{s} \label{algo:bellmanford}}
    \SetKwFor{EmptyLoop}{}{do}{}
    \SetKwFor{parfor}{ParallelForEach}{do}{}
    \SetKw{Break}{break}
    \DontPrintSemicolon
    \myfunc{\upshape$\textsf{Bellman-Ford}(G=(V,E),\ \source\in V)$}{
        $\vname{dist} \gets \mf{Array}.\finit(G.n,\ \infty)$,~$\vname{dist}[\source] \gets 0$ \\
        $\ff \gets \DataTypeA.\finit(G.n,0)$ \tcp*[f]{the current frontier}\\
        $\ff' \gets \DataTypeA.\finit(G.n,0)$ \tcp*[f]{the next frontier}\\
        $\ff.\finsert(\source)$\\
        \For{$\vname{round} \gets 1,2,\ldots,G.n$}{
            \lIf{\upshape$\ff.\fempty()$}{ \Break }
            \hspace{-1mm}
            \EmptyLoop{\upshape$\ff.\fforeach \ v \in \ff$}{
                \parfor{$(u, weight) \in G.N(v)$}{
                    \If{\upshape$\writemin(\vname{dist}[u],\, \vname{dist}[v]{+}weight)$}{
                        $\ff'.\finsert(u)$ \tcp*[f]{add to the next frontier}
                    }
                }
            }
            $\ff.\fclear()$ \tcp*[f]{remove all vertices from the current frontier}\\
            swap $\ff$ and $\ff'$ \tcp*[f]{$\ff$ gets vertices; $\ff'$ becomes empty}\\
        }
        \Return $\vname{dist}[\cdot]$, $\ff.\fempty()$ \tcp*[f]{false if a negative cycle is reached}
    }
\end{algorithm}

%% file: algos/pseudocode_kcore.tex
\begin{algorithm}[t]
\small
    \caption{Parallel \kcore{} using \DataStructure{s} \label{algo:kcore}}
    \SetKwFor{EmptyLoop}{}{do}{}
    \SetKwFor{parfor}{ParallelForEach}{do}{}
    \DontPrintSemicolon
    \myfunc{\upshape\textsf{$k$-core}$(G=(V,E))$}{
        \tcp{induced degree; initialized as the original degree}
        $\vname{induced} \gets \mf{Array}.\finit(G.n,\ v \mapsto d(v))$  \\
        $\fa \gets \DataTypeA.\finit(G.n,\ 1)$  \tcp*[f]{active set: all unpeeled vertices} \\
        $\ff \gets \DataTypeA.\finit(G.n,\ 0)$  \tcp*[f]{the current frontier}\\
        $\ff' \gets \DataTypeA.\finit(G.n,\ 0)$  \tcp*[f]{next frontier}\\
        \While{\upshape$\neg\,\fa.\fempty()$}{
            $k \gets \fa.\freduce(v \mapsto \vname{induced}[v],\min, +\infty)$ \\
            \tcp{extract all vertices with induced degree $k$ to $\ff$}
            \hspace{-1mm}
            \EmptyLoop{\upshape$\fa.\mathrm{\fforeach} \ v \in \fa$}{
                \lIf{\upshape$\vname{induced}[v] = k$}{
                    $\ff.\finsert(v)$
                }
            }
            \While{\upshape$\neg\,\ff.\fempty()$}{
                \hspace{-1mm}
                \EmptyLoop{\upshape$\ff.\fforeach \ v \in \ff$}{
                    $\vname{coreness}[v] \gets k$ \tcp*[f]{assign coreness $k$ to vertex $v$}\\
                    $\fa.\fdelete(v)$ \label{line:kcore-remove}\tcp*[f]{peel $v$ (remove it from $\fa$)}\\
                    \parfor{$u \in G.N(v)$}{
                        \lIf{\upshape$\fa.\fcontains(u)$ \AND $\faadec(\vname{induced}[u], \ 1) = k{+}1$}{
                            $\ff'.\finsert(u)$
                        }
                    }
                }
                $\ff.\fclear()$ \tcp*[f]{remove all vertices from $\ff$}\\
                swap $\ff$ and $\ff'$ \tcp*[f]{$\ff$ gets vertices; $\ff'$ becomes empty}\\
            }
        }
        \Return $\vname{coreness}[\cdot]$
    }
\end{algorithm}

\begin{figure}[t]
  \centering
  \includegraphics[width=\columnwidth]{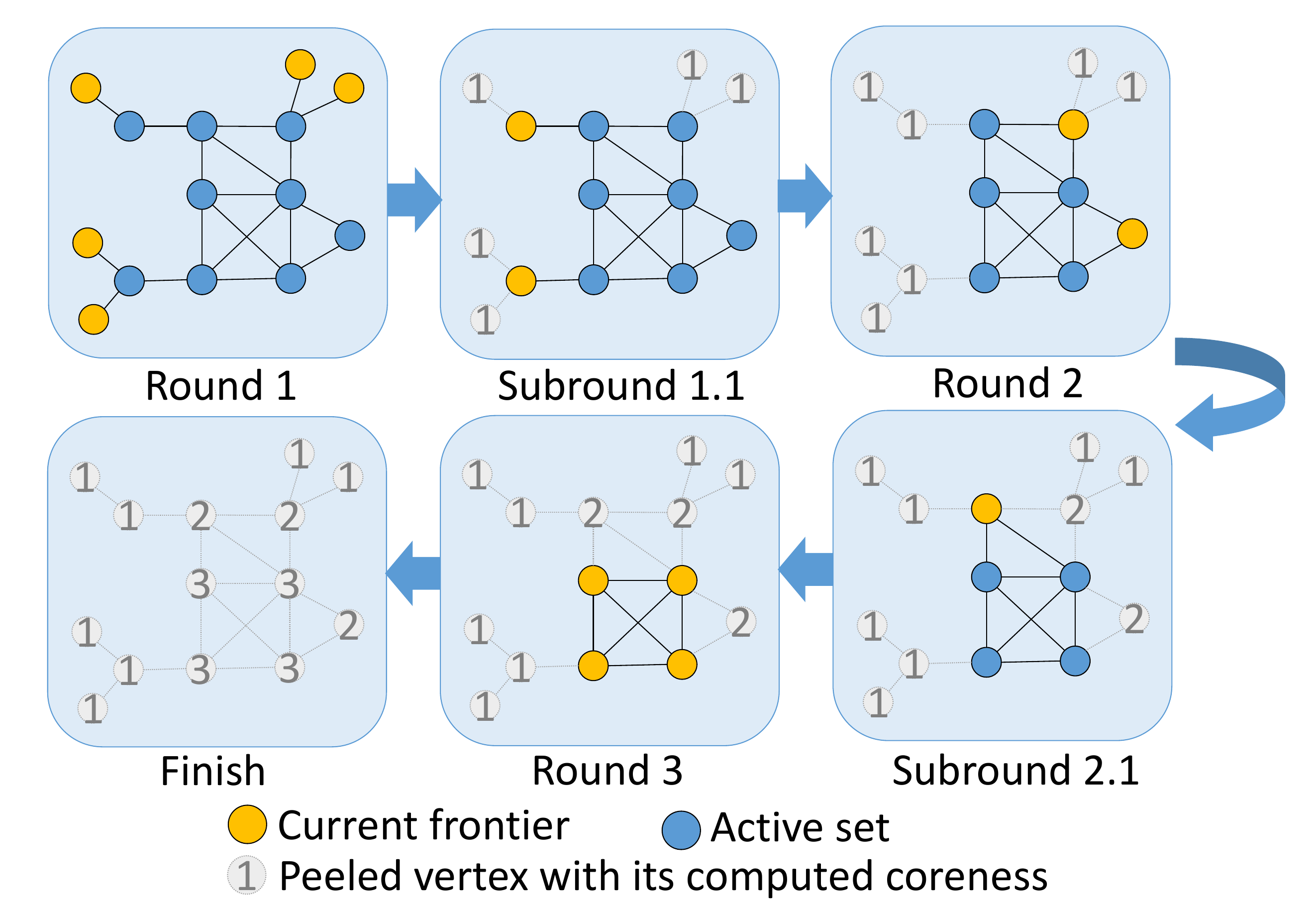}
  \caption{\textbf{An illustration of the peeling process in \kcore{}.}}\label{fig:kcore}
\end{figure}

%% file: theorems/corollary_kcore.tex
\begin{proof}

In the \kcore{} algorithm, $s_i$ in $\fa.\freduce$ is the number of vertices not yet peeled and remaining in $\fa$.

A vertex $v$ remains in the $\fa$ set only as long as its coreness has not yet been determined. 
Specifically, a vertex will be removed from the $\fa$ set in the round where the minimum induced degree $k$ equals $\vname{coreness}[v]$. 
Since $k$ is strictly increasing, a vertex $v$ can stay in the $\fa$ set for at most $(\vname{coreness}[v] + 1)$ rounds.

Summing over all vertices, we have:

$\ \ \ \ s_{\mathit{sum}} = \sum_{\text{number of rounds}} s_i$

$\ \ \ \  = \sum_{v \in V} (\text{number of rounds } v \text{ stays in } \fa)$

$\ \ \ \  \le \sum_{v \in V} (\vname{coreness}[v]+1)$

By the definition of \kcore{}, the coreness of a vertex cannot exceed its initial degree, $\vname{coreness}[v] \le \vname{degree}[v]$. Therefore:

$\ \ \ \ s_{\mathit{sum}} \le \sum_{v \in V} \degree(v) + n = m + n$

Substituting $s_{\mathit{sum}} \le m+n$ into the work bound in \cref{thm:rounds}, and considering that the number of subrounds $\rho$ is no less than the number of rounds, 
we obtain the following bound on the work of $\freduce$ and collecting minimum elements in \cref{algo:kcore}:

$\ \ \ \ O\left((m+n) (1 + \frac{\log (n\rho / (m+n))}{\log \log n})\right)$

$\ \ \ \ \le O\left((m+n) (1 + \frac{\log \rho}{\log \log n})\right)$

Meanwhile, edge updates happen at most $m$ times, and each vertex appears in the frontier at most once, so this part of the work is $O\left(m + n + n \cdot \frac{\log \rho}{\log \log n}\right)$.

Hence, the total work of \cref{algo:kcore} is $O\left((m+n) (1 + \frac{\log \rho}{\log \log n})\right)$,
with $O\left((m+n) \frac{\log \rho}{\log \log n}\right)$ as shorthand (see~\cref{sec:prelim}).
\end{proof}

%% file: algos/pseudocode_bfs.tex
\begin{algorithm}[t]
\small
    \caption{Parallel BFS using \DataStructure{s} \label{algo:bfs}}
    \SetKwFor{EmptyLoop}{}{do}{}
    \SetKwFor{parfor}{ParallelForEach}{do}{}
    \DontPrintSemicolon
    \myfunc{\upshape$\textsf{BFS}(G=(V,E),\ \vname{source}\in V)$}{
        $\vname{dist} \gets \mf{Array}.\finit(G.n,\ \infty)$ \\
        $\vname{mode} \gets \textsc{Forward}$ \\
        $\fa \gets \DataTypeA.\finit(G.n,\ 1)$,~$\fa.\fdelete(\vname{source})$ \\
        $\ff \gets \DataTypeA.\finit(G.n,\ 0)$,~$\ff.\finsert(\vname{source})$ \\
        $\ff' \gets \DataTypeA.\finit(G.n,\ 0)$ \\
        \For{$\vname{round} \gets 0, 1, \ldots$}{
        \eIf(\tcp*[f]{Forward is the default mode}){\upshape$\vname{mode} = \textsc{Forward}$}{
            \lIf{\upshape$\ff.\fempty()$}{\textbf{break}}
            \tcp{Switch mode when frontier is large}
            \If{\upshape$\ff.\freduce(v \mapsto |N(v)|,\ +,\ 0) > \alpha \cdot G.m$\label{line:bfs-switch-1}}{
                \hspace{-1mm}
                \lEmptyLoop{\upshape$\ff.\fforeach \ v \in \ff$}{
                    $\vname{dist}[v] \gets \vname{round}$
                }
                $\ff.\fclear{}()$\\
                $\vname{mode} \gets \textsc{Backward}$ \\
                \textbf{continue}
            }
            \hspace{-1mm}
            \EmptyLoop{\upshape$\ff.\fforeach \ v \in \ff$}{
                $\vname{dist}[v] \gets \vname{round}$\\
                \tcp{out-neighbors for directed graph}
                \parfor{$u \in N(v)$}{
                    \If{\upshape$\fa.\fcontains(u)$\label{line:bfs-concurrency1}}{$\fa.\fdelete(u)$,~$\ff'.\finsert(u)$\label{line:bfs-concurrency2}}
                }
            }
            $\ff.\fclear{}()$\\
            swap $\ff$ and $\ff'$\\
        }(\tcp*[f]{Backward Mode: directional optimization}){
            \hspace{-1mm}
            \EmptyLoop{\upshape$\fa.\fforeach \ v \in \fa$}{
                \tcp{in-neighbors for directed graph}
                \lIf{\upshape$\exists\, u \in N(v): \neg\,\fa.\fcontains(u)$}{$\ff.\finsert(v)$}
            }
            \lIf{\upshape$\ff.\fempty()$}{\textbf{break}}
            \tcp{Switch mode when frontier is sparse}
            \If{\upshape$\ff.\freduce(v \mapsto 1,\ +,\ 0) < \beta \cdot G.n$\label{line:bfs-switch-2}}{ 
                \hspace{-1mm}
                \lEmptyLoop{\upshape$\ff.\fforeach \ v \in \ff$}{
                    $\fa.\fdelete(v)$
                }
                $\vname{mode} \gets \textsc{Forward}$,~$round \gets round-1$ \\
                \textbf{continue}
            }
            \hspace{-1mm}
            \EmptyLoop{\upshape$\ff.\fforeach \ v \in \ff$}{
                $\fa.\fdelete(v)$,~$\vname{dist}[v] \gets \vname{round}$
            }
            $\ff.\fclear{}()$\\
        }
        }
        \Return $\vname{dist}[\cdot]$
    }
\end{algorithm}

%% file: algos/pseudocode_coloring.tex
\begin{algorithm}[t]
\small
    \caption{Parallel Largest-Degree-First (LF) Graph Coloring Algorithm using \DataStructure{s} \label{algo:coloring}}
    \SetKwFor{EmptyLoop}{}{do}{}
    \SetKwFor{parfor}{ParallelForEach}{do}{}
    \DontPrintSemicolon
    \myfunc{\upshape$\textsf{Coloring}(G=(V,E))$}{
        $\vname{color} \gets \mf{Array}.\finit(G.n,\ \infty)$ \\
        $\vname{perm} \gets $ a random permutation of $[0, 1, ..., G.n-1]$ \\
        $\vname{priority} \gets \mf{Array}.\finit(G.n,\ \infty)$ \\
        $\ff \gets \DataTypeA.\finit(G.n,\ 0)$ \\
        $\ff' \gets \DataTypeA.\finit(G.n,\ 0)$ \\
        \parfor{$v\in V$}{
            \tcp{$d(v)$ represents the degree of vertex $v$}
            $\vname{priority}[v]\gets \left|u \in G.N(v) : (d(u) > d(v))\right.$\\
            $\ \ \ \ \ \ \ \ \ \ \ \ \ \ \ \ \ \ \ \ \ \left.\lor (d(u) = d(v) \land \vname{perm}[u] < \vname{perm}[v])\right|$ \\
            \lIf{$\vname{priority}[v] = 0$}{$\ff.\finsert(v)$}
        }
        \While{\upshape$\neg\,\ff.\fempty()$}{
            \hspace{-1mm}
            \EmptyLoop{\upshape$\ff.\fforeach \ v \in \ff$}{
                $\vname{used} \gets \{\vname{color}[u] : u \in N(v),\, \vname{color}[u] \ne \infty\}$\label{line:coloring-color1}\\
                $\vname{color}[v] \gets \min\{c \ge 0 : c \notin \vname{used}\}$\label{line:coloring-color2}\\
                \parfor{$u \in N(v)$ with $\vname{color}[u] = \infty$}{
                    \lIf{\upshape$\faadec(\vname{priority}[u],\ 1) = 1$}{$\ff'.\finsert(u)$}
                }
            }
            $\ff.\fclear{}()$\\
            swap $\ff$ and $\ff'$\\
        }
        \Return $\vname{color}[\cdot]$
    }
\end{algorithm}

%% file: algos/pseudocode_wBFS.tex
\begin{algorithm}[h]
\small

    \caption{Parallel weighted-BFS (wBFS) using {\DataStructure}s with \DataTypeA and \DataTypeB interface, respectively \label{algo:wbfs1}}
    \SetKwFor{EmptyLoop}{}{do}{}
    \SetKwFor{parfor}{ParallelForEach}{do}{}
    \DontPrintSemicolon
    \myfunc(\tcp*[f]{\DataTypeB version}){\upshape$\textsf{wBFS}(G=(V,E),\ \vname{source}\in V)$}{
        $\vname{dist} \gets \mf{Array}.\finit(G.n,\ +\infty)$,~$\vname{dist}[\vname{source}] \gets 0$ \\
        $\fa \gets \DataTypeB.\finit(G.n)$,~$\fa.\fdecval(\vname{source},\ 0)$ \\
        $\ff \gets \DataTypeA.\finit(G.n,\ 0)$\\
        \While{$true$}{
            \tcp{extract settled vertices to the frontier}
            \hspace{-1mm}
            \lEmptyLoop{\upshape$\fa.\fextract \ v\in \fa$\label{line:wbfs-extract}}{
                $\ff.\finsert(v)$
            }
            \lIf(\tcp*[f]{algorithm complete}){\upshape$\ff.\fempty()$}{\Return $\vname{dist}[\cdot]$}
            \hspace{-1mm}
            \EmptyLoop{\upshape$\ff.\fforeach \ v\in \ff$}{
               \parfor{$(u,w) \in G.N(v)$}{
                    \If{\upshape$\writemin(\vname{dist}[u],\ \vname{dist}[v] + w)$}{
                        $\fa.\fdecval(u,\ \vname{dist}[v] + w)$
                    }
                }
            }
            $\ff.\fclear{}()$\\
        }
    }

    \medskip

    \SetKwFor{EmptyLoop}{}{do}{}
    \SetKwFor{parfor}{ParallelForEach}{do}{}
    \DontPrintSemicolon
    \myfunc(\tcp*[f]{\DataTypeA version}){\upshape$\textsf{wBFS}(G=(V,E),\ \vname{source}\in V)$}{
        $\vname{dist} \gets \mf{Array}.\finit(G.n,\ +\infty)$,~$\vname{dist}[\vname{source}] \gets 0$ \\
        $\fa \gets \DataTypeA.\finit(G.n,\ 0)$,~$\fa.\finsert(\vname{source})$ \\
        $\ff \gets \DataTypeA.\finit(G.n,\ 0)$ \\
        $round \gets 0$ \\
        \While(\tcp*[f]{process vertices with dist $=\vname{round}$}){\upshape$\neg~\fa.\fempty()$}{
            \hspace{-1mm}
            \EmptyLoop{\upshape$\fa.\fforeach \ v \in \fa$}{
                \tcp{vertices with dist $=\vname{round}$ are settled}
                \lIf(){$\vname{dist}[v] = \vname{round}$}{
                    $\ff.\finsert(v)$
                }
            }
            \tcp{remove settled vertices from $\fa$}
            \hspace{-1mm}
            \lEmptyLoop{\upshape$\ff.\fforeach \ v \in \ff$}{
              $\fa.\fdelete(v)$
            }
            \hspace{-1mm}
            \EmptyLoop(\tcp*[f]{relax their neighbors}){\upshape$\ff.\fforeach \ v \in \ff$}{
                \parfor{$(u,w) \in G.N(v)$}{
                    \tcp{if the relaxation is successful, add $u$ to $\fa$}
                    \lIf{\upshape$\writemin(\vname{dist}[u],\ \vname{dist}[v] + w)$}{
                        $\fa.\finsert(u)$
                    }
                }
            }
            $\ff.\fclear{}()$\\
            $round \gets round+1$ \\
        }
        \Return $\vname{dist}[\cdot]$
    }
    
\end{algorithm}

%% file: main/05_experiment.tex
\section{Experiments}\label{sec:exp}

\myparagraph{Setup.} We run experiments on a 96-core machine (192 hyper-threads) equipped with four 2.1 GHz Intel Xeon Gold 6252 CPUs, 
each with a 36 MB L3 cache and 1.5 TB of main memory. 
For all parallel tests, we used \texttt{numactl -i all} to interleave memory across all CPUs. 
Our code is in C++ and uses ParlayLib~\cite{blelloch2020parlaylib} for a parallel scheduler.
All implementations are compiled using GCC~15 with \texttt{-O3 -march=native -std=c++20 -pthread} (plus \texttt{-fopenmp} for the OpenMP-based baselines), and the main experiments use all 192 hyper-threads. 
For algorithms with sources, we choose five random sources from the largest connected component, run the algorithms on each source twice (the first time for warm-up) and report the average of the second runs. 
For all other algorithms, we run them six times and report the average of the last five.

To make the experimental inputs reproducible, we deterministically assign each edge $(u,v)$ in a weighted graph $G$ the weight $1+\texttt{hash64}(2^{32}\cdot\min(u,v)+\max(u,v))\bmod (\lfloor\log_2 G.n\rfloor+1)$, so that both directions of an edge receive the same weight.
For single-source problems, the sources are the first five candidates $\texttt{hash32}(i)\bmod G.n$ ($i=0,1,2,\ldots$) in the largest connected component of $G$.
Here $\texttt{hash32}$ and $\texttt{hash64}$ are the hash functions from ParlayLib~\cite{blelloch2020parlaylib}.
We verified that all implementations produce bitwise-identical outputs. 

\myparagraph{Datasets.} 
We use a diverse collection of graphs, including both low-diameter (social networks, web graphs) and large-diameter graphs (road networks, $k$-NN, and synthetic mesh-like graphs). 
For parallel graph algorithms, a lower diameter usually means fewer rounds, and thus better parallelism.
Since $k$-core and coloring are defined on undirected graphs, we symmetrize all graphs to make them consistent across applications. 
The symmetrized graph information is listed in \cref{table:fulltable}.

\input{experiments/main/main_figures.tex}

\input{tables/graph_full_info.tex}

\subsection{Overall Performance}
\label{sec:exp:app}

\myparagraph{Settings and Baselines.}
We implement the graph algorithms in \cref{sec:applications}, integrating \ourtree{} into state-of-the-art approaches. 
To isolate performance gains from the data structure, we carefully align our algorithmic techniques~\cite{totcode} with the baselines. 
We then compare the end-to-end performance of our implementation against these baselines. 
We note that all studied applications are fundamental graph problems with extensive existing work. 
For implementation and comparison, our techniques and baselines mainly come from two graph libraries: GBBS~\cite{gbbs2021} and PASGAL~\cite{dong2024pasgal},
with their algorithms proposed in several relevant papers~\cite{gbbs2021,dhulipala2017,dong2024pasgal,wang2023parallel,liu2025parallel,shun2013ligra}.

For all five algorithms, we include the implementations from GBBS~\cite{gbbs2021} as baselines, 
with the relevant techniques proposed in~\cite{dhulipala2017,shun2013ligra}. 
GBBS is a well-known graph processing library, providing state-of-the-art performance for a wide range of graph algorithms. 
It uses a two-pass traversal approach (see \cref{sec:discussion}) to maintain vertex sets.

We also include $k$-core and BFS from PASGAL~\cite{dong2024pasgal}, a more recent library that improved upon GBBS in these two problems~\cite{dong2024pasgal,liu2025parallel}. 
PASGAL uses \hashbag{s} (see \cref{sec:discussion}) to maintain frontiers, and vertical granularity control (VGC) to optimize performance on large-diameter graphs~\cite{wang2023parallel}. 
Its $k$-core algorithm includes sampling for large-degree nodes~\cite{liu2025parallel}. We adopt these optimizations in our implementations. 

We additionally include GAPBS~\cite{beamer2015gap} (the implementation released by the authors of direction-optimizing BFS~\cite{Beamer12}) and Galois~\cite{nguyen2013lightweight} as baselines in BFS, ParK~\cite{dasari2014park} and PKC~\cite{kabir2017parallel} as baselines in $k$-core.

\myparagraph{Performance Analysis.} We now present an end-to-end performance comparison with existing work. 
For clarity, we present the relative performance of all baselines in \cref{fig:main_fa},
with a summary of the geometric mean of the running times in each category.
The raw running times are provided in \cref{tab:rawdata_main} in the appendix.

Across all tests, our implementations achieve a significant advantage over the baselines. 
Our average running time across all graphs is at least \emph{1.8$\times{}$ faster than all the baselines for each graph problem}. 
Among the 100 tested algorithm-graph combinations, our solution is faster than every baseline in 98 cases.
Our solution is at least 2$\times$ faster than the \emph{strongest baseline} in more than 40 percent of the test instances (41/100),
and is at least 1.5$\times$ faster in more than 70 percent of them (72/100). 
These results provide strong evidence for the robust and superior performance of the \ourtree{}, indicating that
using it can lead to improved performance on these parallel graph problems over the state of the art.

\subsection{In-Depth Performance Study}

\begin{figure}[t]
  \centering
  \includegraphics[width=\columnwidth]{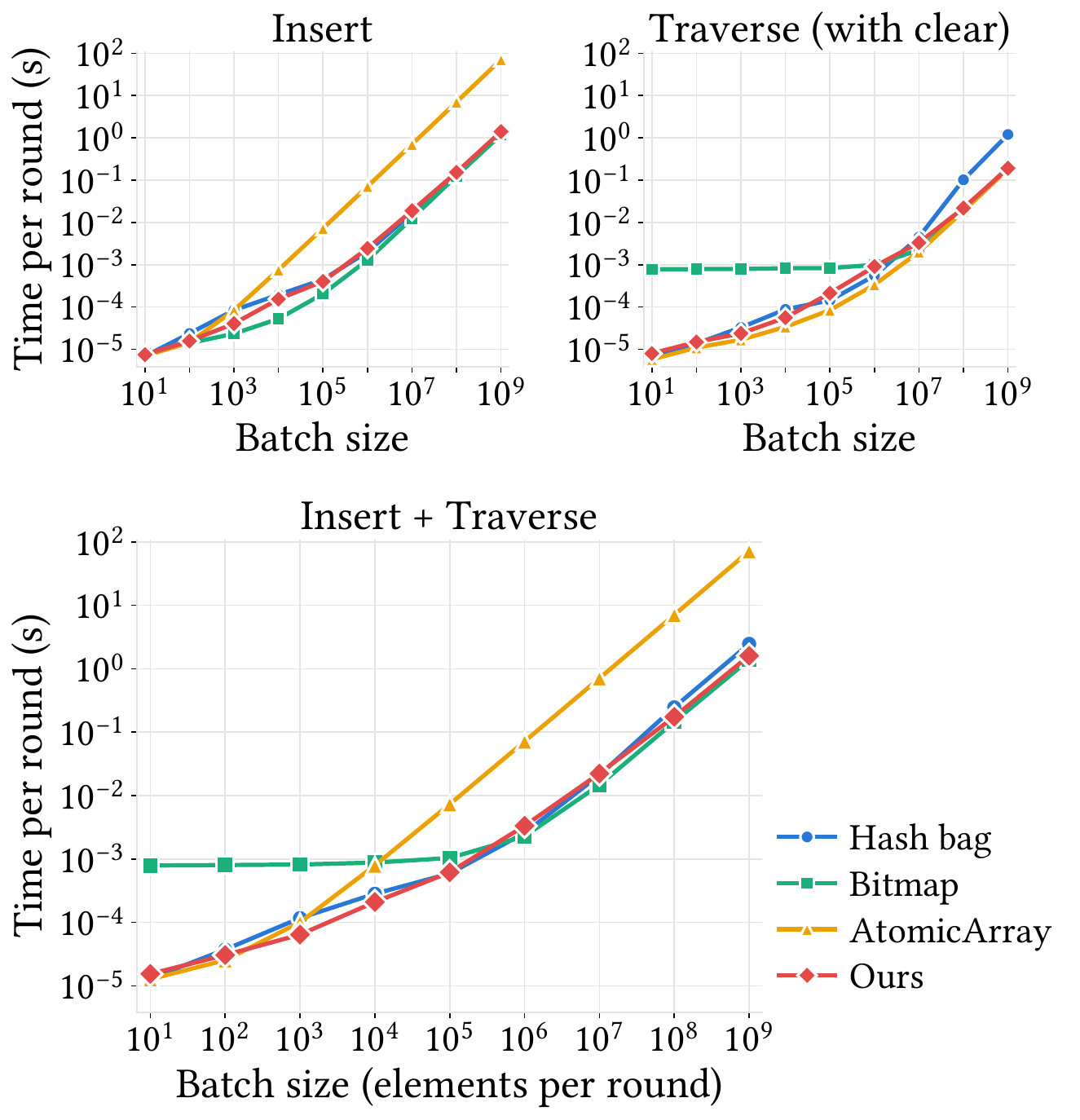}
  \caption{\textbf{Microbenchmark of \ourtree{} (ours), Bitmap, AtomicArray, and \hashbag{}.}
  $10^9$ random insertions are split into rounds of varying batch sizes; each round inserts a batch, traverses the set, and clears it.
  We report the average time per round for insertion, traversal (including clearing), and their sum. Lower is better.}
  \label{fig:microbench}
\end{figure}

\myparagraph{Comparison to Simple Representations.}\label{sec:exp:simple}
We compare the \ourtree{} with three simple representations outside the application context: a flat \emph{Bitmap}, an \emph{AtomicArray} (a list of indices with atomic append), and the \hashbag{}.
We perform $10^9$ insertions at distinct random indices in a universe of size $10^9$, divided into rounds with batch sizes from $10$ to $10^9$.
Each round inserts one batch, traverses the set, and clears it.
\Cref{fig:microbench} shows the results.
Bitmap has the fastest insertions and AtomicArray the fastest traversals, but neither performs well across all batch sizes:
AtomicArray suffers from contention on its shared counter once the batch size reaches about $10^3$, and Bitmap must scan the entire bit vector regardless of the set size, which dominates for sparse sets.
Altogether, the \ourtree{} achieves the best overall performance: it is the fastest for intermediate batch sizes (around $10^2$--$10^5$) and remains close to the fastest alternative at both extremes.

\label{exp-compare-hashbag}
\myparagraph{Comparison to the \hashbag.} 
We compare \ourtree{} to its counterpart, the \hashbag~\cite{wang2023parallel}, which provides a similar interface for frontier processing. 
We substitute \ourtree{} with the PASGAL \hashbag in four problems: Bellman-Ford, $k$-core, BFS, and coloring. 
To isolate data structure effects, we disable most optimizations~\cite{totcode} (e.g., VGC) in both versions. 
\cref{tab:main_table_hashbag} summarizes the results; the per-graph relative running times (\cref{fig:hashbag_comparison}) and the raw data (\cref{tab:rawdata_hashbag}) are given in the appendix.

\input{experiments/main/main_table_hashbag.tex}

By replacing only the frontier data structure~\cite{totcode}, \DataStructure{s} outperform \hashbag{s} by 2.52$\times$ in BFS, 1.69$\times$ in $k$-core, 1.59$\times$ in coloring, and 2.09$\times$ in Bellman-Ford. 
Across all problem-graph combinations, \DataStructure{s} consistently outperform \hashbag{s}. 
To understand the performance gains, we select 4 representative graph-problem combinations. \cref{fig:rounds} plots the performance of \ourtree{s} and \hashbag{s} across varying frontier sizes. Each point $(x,y)$ represents a single round processing a frontier of size $x$ with running time $y$ in this round.  
Solid lines indicate the corresponding linear fits.

\input{figures/rounds.tex}

Across all figures, the fitted lines for \DataStructure{s} exhibit smaller slopes than \hashbag{s}, indicating lower overhead in practice:
While \DataStructure{s} are slower on small frontiers for certain graphs, this regime contributes negligibly to the total end-to-end runtime. 
Conversely, large frontiers dominate the total execution time, a regime where the \DataStructure{} maintains a clear advantage. 
These observations help explain why \DataStructure{} outperforms \hashbag{} overall in large-scale graph processing, as shown in \cref{tab:main_table_hashbag}.

\myparagraph{\DataTypeA{} vs. \DataTypeB{} for wBFS.} 
We tested two wBFS variants using \DataTypeA{} and \DataTypeB{}. 
In almost all cases, both outperform the baseline, with \DataTypeB{} having a slight advantage. 
This aligns with our theoretical analysis (\cref{col:wbfs}). 
While \DataTypeB{} accelerates extraction, its updates become slower. 
Hence, \DataTypeB{} remains preferable in this use case, despite the small performance gap.

\input{experiments/speedup/speedup.tex}
\myparagraph{Self-relative Speedup.} 
\cref{fig:speedup} presents the self-relative speedups for \ourtree{}, PASGAL and GBBS. 
GBBS shows relatively low BFS speedups on road networks due to their inherent sparsity, 
while our code and PASGAL using VGC are more resistant to this. 
Otherwise, the self-relative speedups of all algorithms remain consistent. 
This demonstrates that our scalability matches that of state-of-the-art parallel graph libraries,
and further \emph{suggests} that our performance gains stem primarily from smaller constant factors.

%% file: experiments/main/main_figures.tex
\begin{figure*}[t]
  \centering
  \includegraphics[width=\linewidth]{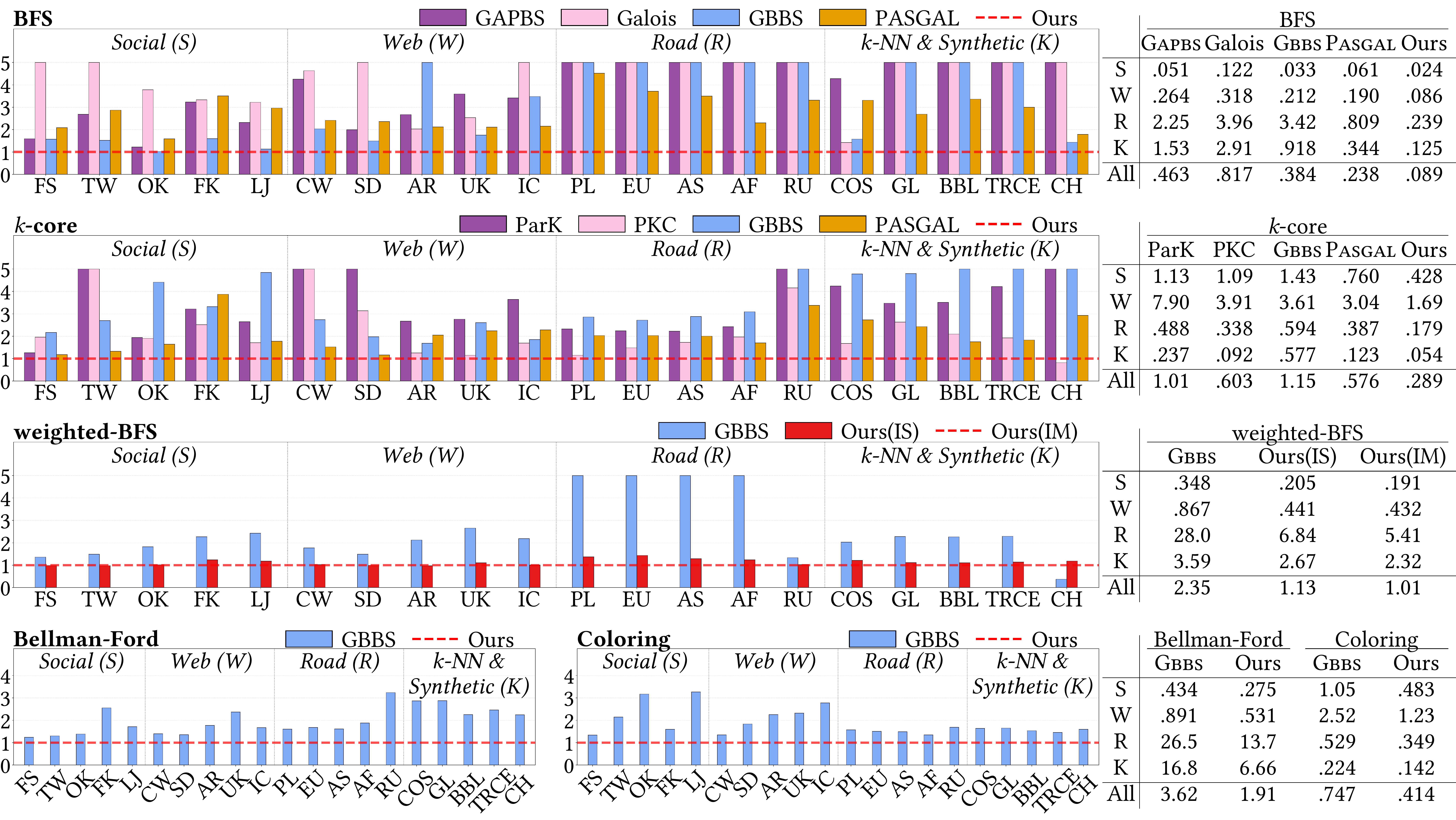}
  \caption{\small 
  \textbf{Relative running time for all baselines (introduced in \cref{sec:exp:app}) over our implementation (red dotted line at 1). 
    }
    Values above 1 favor our implementation, with higher bars indicating larger speedups (values above 5 are clipped); values below 1 favor the baseline. 
    In the table, each number is the geometric mean of all graphs in the same category. ``All'': geometric mean among all tested graphs. 
    Ours (IS) and Ours (IM) use \DataTypeIS and \DataTypeIM, respectively.\vspace{.75em}
    }
  \label{fig:main_fa}
\end{figure*}

%% file: tables/graph_full_info.tex
\begin{table}[t]
\small
  \centering
    \begin{tabular}{@{}c@{\hspace{0pt}}c@{\hspace{8pt}}r@{\hspace{8pt}}r@{\hspace{8pt}}l@{}}
    \multicolumn{1}{c}{\textbf{Category}} & \textbf{Graph} & \multicolumn{1}{c}{$\boldsymbol{n}$} & \multicolumn{1}{c}{$\boldsymbol{m}$} & \multicolumn{1}{c}{\textbf{Notes}} \\
    \midrule
    \midrule
    \multicolumn{1}{c}{\multirow{5}{*}{\rotatebox{90}{Social}}}
        & FS    & 65.6M & 3.6B       & Friendster~\cite{yang2015defining} \\
        & TW    & 41.7M & 2.4B       & Twitter~\cite{kwak2010twitter} \\
        & OK    & 3.1M  & 234.4M     & com-orkut~\cite{yang2015defining} \\
        & FK    & 59.2M & 185.0M     & socfb-konect~\cite{nr,traud2012social,red2011comparing} \\
        & LJ    & 4.8M  & 85.7M      & soc-LiveJournal1~\cite{backstrom2006group} \\
    \midrule
    \multicolumn{1}{c}{\multirow{5}{*}{\rotatebox{90}{Web}}}
        & CW    & 978.4M & 74.7B     & ClueWeb~\cite{webgraph} \\
        & SD    & 89.2M & 3.9B       & sd-arc~\cite{webgraph} \\
        & AR    & 22.7M & 1.1B       & arabic~\cite{nr} \\
        & UK    & 18.5M & 523.6M     & uk-2002~\cite{webgraph} \\
        & IC    & 7.4M  & 302.0M     & indochina~\cite{webgraph} \\
    \midrule
    \multicolumn{1}{c}{\multirow{5}{*}{\rotatebox{90}{Road}}}
        & PL    & 372.4M & 953.6M    & Planet road network~\cite{roadgraph} \\
        & EU    & 130.7M & 332.6M    & OSM Europe~\cite{roadgraph} \\
        & AS    & 95.7M & 243.6M     & OSM Asia~\cite{roadgraph} \\
        & AF    & 33.5M & 88.9M      & OSM Africa~\cite{roadgraph} \\
        & RU    & 23.9M & 57.7M      & RoadUSA~\cite{roadgraph} \\
    \midrule
    \multicolumn{1}{c}{\multirow{5}{*}{\rotatebox{90}{\shortstack{$k$-NN \& \\ Synthetic}}}}
        & COS   & 321.1M & 1.96B     & Cosmo50~\cite{cosmo50,wang2021geograph} \\
        & GL    & 24.9M  & 309.7M    & GeoLife~\cite{geolife,wang2021geograph} \\
        & BBL   & 21.2M  & 63.6M     & hugebubbles-00020~\cite{nr} \\
        & TRCE  & 16.0M  & 48.0M     & hugetrace-00020~\cite{nr} \\
        & CH    & 4.2M   & 29.7M     & Chem~\cite{fonollosa2015reservoir,wang2021geograph} \\
    \bottomrule
    \end{tabular}%

    \vspace{1em}
    \caption{\textbf{Information of all tested graphs.}
    $n$: number of vertices. 
    $m$: number of directed CSR edge entries.
    OSM: OpenStreetMap.
    }
    \label{table:fulltable}
\end{table}

%% file: experiments/main/main_table_hashbag.tex
\begin{table}
\fontsize{8pt}{9.6pt}\selectfont
\setlength{\tabcolsep}{0pt}
\begin{tabularx}{\columnwidth}{@{}l|*{3}{>{\raggedright\arraybackslash}X>{\raggedleft\arraybackslash}X|}>{\raggedright\arraybackslash}X>{\raggedleft\arraybackslash}X@{}}
\toprule
& \multicolumn{2}{c}{BFS} & \multicolumn{2}{c}{$k$-core} & \multicolumn{2}{c}{Coloring} & \multicolumn{2}{c}{Bellman-Ford} \\
 & \makebox[0pt][l]{\hspace{3pt}\hashbag} & Ours\hspace*{3pt} & \makebox[0pt][l]{\hspace{3pt}\hashbag} & Ours\hspace*{3pt} & \makebox[0pt][l]{\hspace{3pt}\hashbag} & Ours\hspace*{3pt} & \makebox[0pt][l]{\hspace{3pt}\hashbag} & Ours\hspace*{3pt} \\
 \midrule
Social\hspace*{1pt} & \hspace{8pt}.076 & .025\hspace*{8pt} & \hspace{8pt}.813 & .574\hspace*{8pt} & \hspace{8pt}.656 & .483\hspace*{8pt} & \hspace{8pt}.503 & .275\hspace*{8pt} \\
Web & \hspace{8pt}.386 & .087\hspace*{8pt} & \hspace{8pt}3.05 & 2.75\hspace*{8pt} & \hspace{8pt}1.70 & 1.23\hspace*{8pt} & \hspace{8pt}1.33 & .531\hspace*{8pt} \\
Road & \hspace{8pt}1.80 & 1.12\hspace*{8pt} & \hspace{8pt}.494 & .242\hspace*{8pt} & \hspace{8pt}.664 & .349\hspace*{8pt} & \hspace{8pt}23.7 & 13.7\hspace*{8pt} \\
$k$-NN & \hspace{8pt}1.04 & .563\hspace*{8pt} & \hspace{8pt}.323 & .128\hspace*{8pt} & \hspace{8pt}.254 & .142\hspace*{8pt} & \hspace{8pt}16.0 & 6.66\hspace*{8pt} \\
\midrule
Avg. & \hspace{8pt}.484 & .192\hspace*{8pt} & \hspace{8pt}.793 & .470\hspace*{8pt} & \hspace{8pt}.659 & .414\hspace*{8pt} & \hspace{8pt}3.99 & 1.91\hspace*{8pt} \\
\bottomrule
\end{tabularx}
\medskip
\captionof{table}{
\textbf{Running time (seconds) comparison with \emph{\hashbag}. Lower is better.}
``\hashbag'': use \hashbag in PASGAL to replace \ourtree{} in our code. 
``$k$-NN'': $k$-NN and synthetic graphs. 
Raw data of per-graph running time is in \cref{tab:rawdata_hashbag}.
Each number is the geometric mean of all graphs in the same category.
``Avg.'': geometric mean among all tested graphs.
}
\label{tab:main_table_hashbag}
\end{table}

%% file: figures/rounds.tex
\begin{figure}[t]
\centering
\includegraphics[width=\columnwidth]{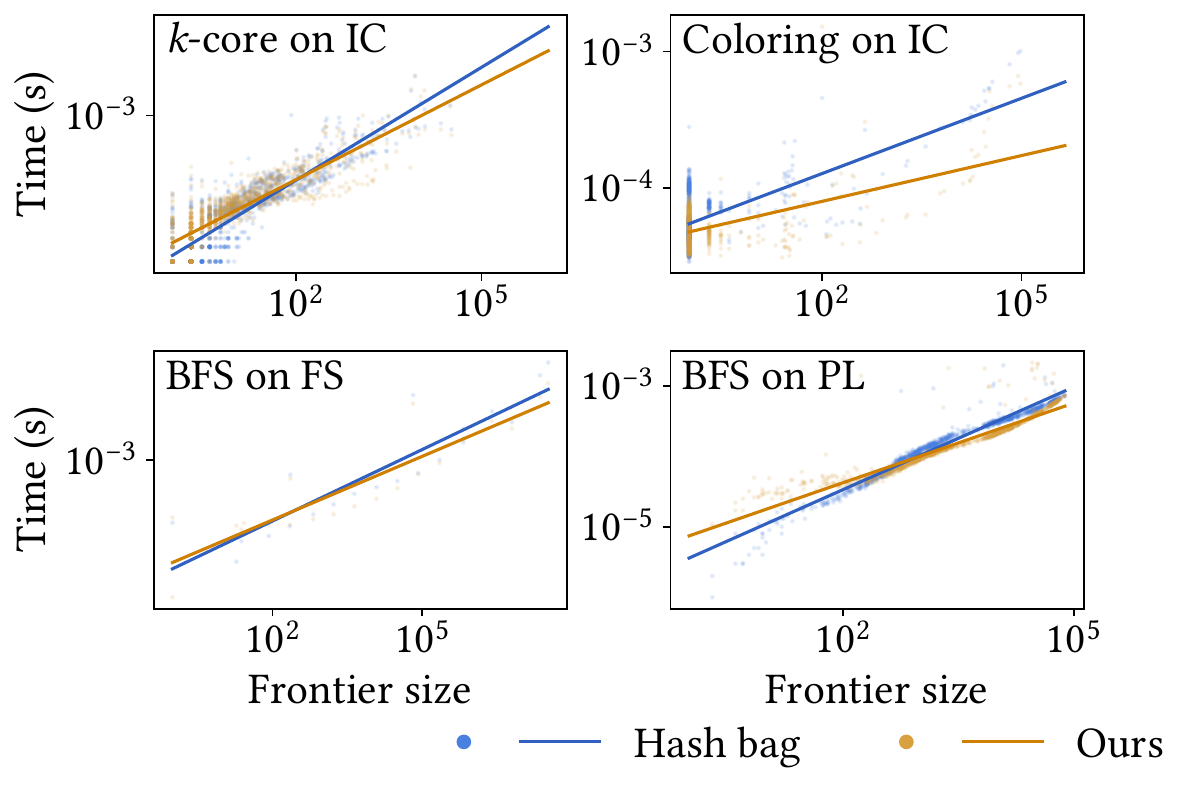}
\caption{\textbf{Per-round time comparison of \hashbag{} and \DataStructure{} on \kcore{}, graph coloring, and BFS on selected graphs}. 
Each point $(x,y)$ represents one round, where $x$ is the frontier size and $y$ the time in seconds. Lines show log--log linear regression fits.
}
\label{fig:rounds}
\end{figure}

%% file: experiments/speedup/speedup.tex
\begin{figure}
\includegraphics[width=\columnwidth]{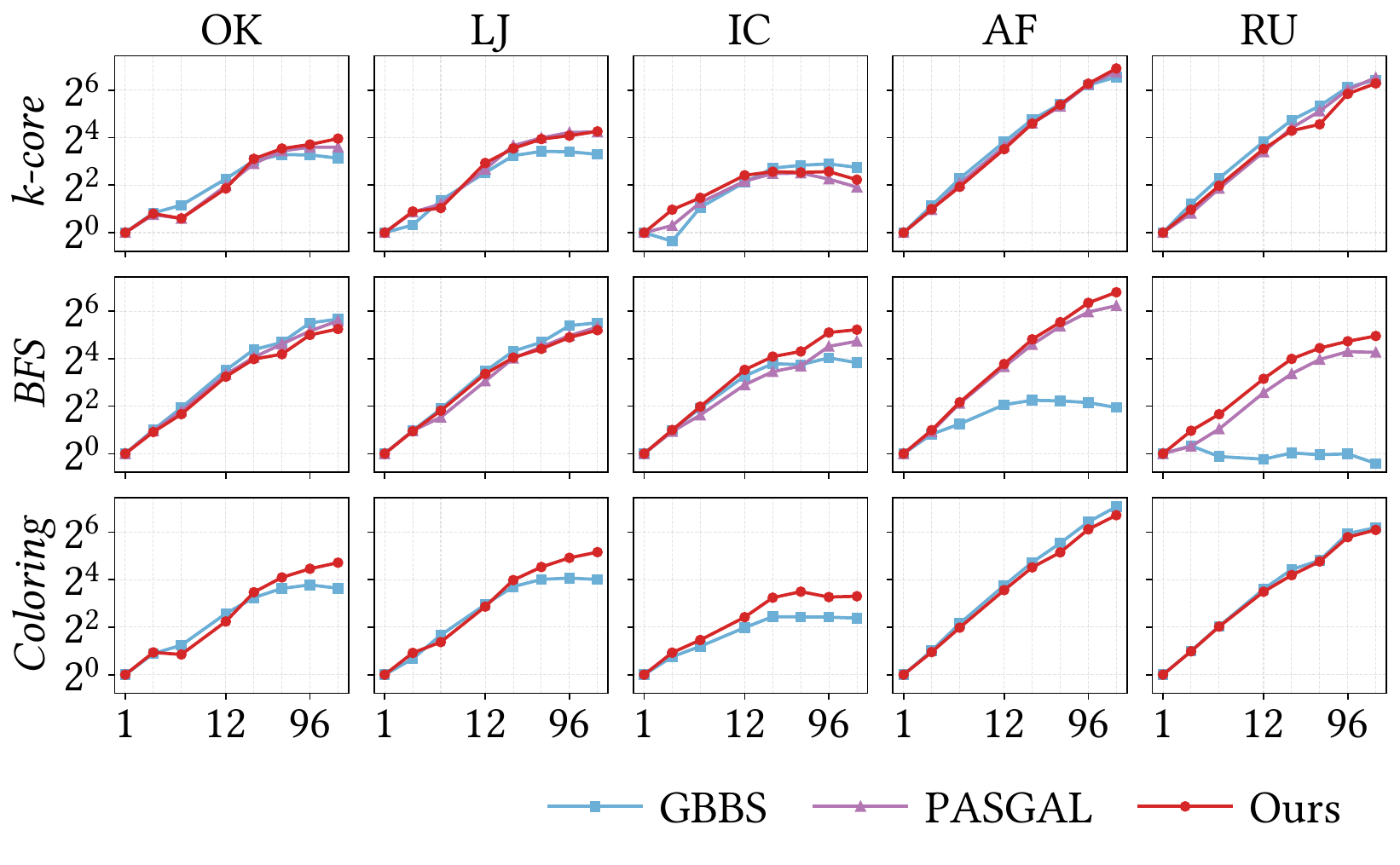}
\captionof{figure}{\textbf{Self-relative speedup (log-log scale) of our code, GBBS \cite{gbbs2021} and PASGAL~\cite{dong2024pasgal}. 
Higher is better.}  The $x$-axis shows 1, 2, 4, 12, 24, 48, and 96 cores, plus 96 cores with hyperthreading.
}
\label{fig:speedup}
\end{figure}

%% file: main/06_relatedwork.tex
\section{Related Work}
Many data structures have been proposed for frontiers in graph processing~\cite{Hong2017multigraph,wang2023parallel,shun2013ligra}. 
We reviewed some of the most relevant ones in \cref{sec:discussion}. 
There are other data structures for \DataTypeA{}, such as the vEB tree~\cite{van1977preserving}. 
Our work does not focus on improving the theoretical bounds, but on achieving good practical performance with low theoretical overhead.

From a broader perspective, many graph-processing systems are frontier-based, including shared-memory frameworks such as Ligra~\cite{shun2013ligra},
GBBS~\cite{gbbs2021}, Galois~\cite{Nguyen2014}, and GraphIt~\cite{zhang2018graphit,zhang2020optimizing}, and GPU ones such as Merrill et al.'s framework~\cite{merrill2012scalable},
Gunrock~\cite{wang2016gunrock}, and GraphBLAST~\cite{yang2022graphblast}. 
Most of them use an approach similar to that of GBBS with some platform-specific optimizations.
Our work is complementary
to these systems: rather than proposing a new programming model or graph algorithm, 
the \ourtree{} provides an efficient CPU data structure for dynamically maintaining and traversing
vertex subsets under parallel updates.
While \ourtree{} shares a similar representation to sequential hierarchical bit vectors~\cite{fraenkel1986improved,glenn2008investigation} and Rust's \texttt{hibitset}~\cite{hibitset}, its data-structural contribution lies in the \emph{parallel} interfaces and algorithms with refined work bounds, rather than in the hierarchical representation itself.

Using tournament trees for parallel priority queues has been explored~\cite{dong2021efficient,wang2023fast,gu2023parallel}. 
Our \DataTypeB{} idea is similar to~\cite{dong2021efficient}. 
The authors of \cite{dong2021efficient} hypothesized that the tournament tree is less practical than array-based structures for their SSSP algorithm. 
Our work demonstrates that the \ourtree{} is still effective in wBFS. 

All graph problems in this paper have been extensively studied (e.g.~\cite{shun2013ligra,gbbs2021,schidler2023sat,el2023b,kabir2017parallel,liu2025parallel,welsh1967upper,jones1993parallel,hasenplaugh2014ordering,lv2012compression,green2021inverse,xiao2025case,Beamer12,zhang2020optimizing,chen2017efficient,merrill2012scalable,merrill2015high,dhulipala2017,gregor2005lifting,alabandi2020increasing,liu2015enterprise,rai2019hybrid,osama2019graph,peng2018graphphi}).
Our work does not provide new algorithms for these problems, but rather a more efficient data structure to facilitate them.
We believe many of them can also benefit from \ourtree{}. 

%% file: main/07_conclusion.tex
\section{Conclusion}\label{sec:conclusion}

We propose the \ourtree{}, a lightweight parallel data structure for maintaining vertex subsets in frontier-based graph algorithms. 
It combines the advantages of traditional sparse and dense representations. 
The design uses bit vectors to provide easy dynamic updates (similar to the dense mode), and a tree hierarchy for efficient traversal (similar to the sparse mode). 
\ourtree{s} provide interfaces for \DataTypeA{} and \DataTypeB{}, which are readily applicable to various parallel algorithms as shown in this paper. 

As examples, we apply \ourtree{}s to five fundamental graph problems: BFS, \kcore{}, graph coloring, SSSP using Bellman-Ford, and weighted-BFS (wBFS), and analyze their cost bounds. 
All of them achieve the same or nearly the same (up to a logarithmic overhead) work bounds as existing work. 

Our experiments show that \ourtree{}s are highly efficient in practice. 
Across a broad set of tested graphs, our approach outperforms state-of-the-art baselines, achieving an average speedup of at least 1.8$\times$ over the best baseline in all tested problems. 
These results demonstrate the effectiveness of \ourtree{}s for high-performance parallel algorithms.
We believe that \ourtree{}s will serve as a highly promising foundation for future graph algorithms and frameworks.

\myparagraph{Acknowledgments.} We used AI to assist with figure generation, code debugging, and language polishing.
We reviewed all AI-assisted material and take full responsibility for them.

%% file: main/99_appendix.tex
\appendix

\section{Raw Experimental Running Times}
\label{sec:rawdata}

Here we present the raw running times of our algorithms and all baselines in \cref{tab:rawdata_main} and \cref{tab:rawdata_hashbag}.
\Cref{fig:hashbag_comparison} visualizes the data in \cref{tab:rawdata_hashbag}.

\input{experiments/main/rawdata.tex}

\input{experiments/main/hashbag_figures.tex}

%% file: experiments/main/rawdata.tex
\begin{table*}[!h]
\small
\renewcommand{\arraystretch}{0.9}
\begin{tabular}{llcccccccc}
\toprule
\multicolumn{2}{c}{} & \multicolumn{5}{c}{BFS} & \multicolumn{3}{c}{weighted-BFS} \\[2pt]
\cmidrule(lr){3-7}
\cmidrule(lr){8-10}
 &  & GBBS & PASGAL & GAPBS & Galois & Ours & GBBS & Ours(IS) & Ours(IM) \\
\midrule
\multirow{6}{*}{\rotatebox[origin=c]{90}{Social}} & FS & 0.164 & 0.218 & 0.166 & 0.903 & \textbf{0.104} & 1.73 & \textbf{1.26} & 1.27 \\
 & TW & 0.0676 & 0.128 & 0.120 & 0.396 & \textbf{0.0445} & 0.843 & \textbf{0.554} & 0.566 \\
 & OK & 0.00743 & 0.0116 & 0.00894 & 0.0276 & \textbf{0.00730} & 0.0812 & 0.0455 & \textbf{0.0446} \\
 & FK & 0.0550 & 0.121 & 0.111 & 0.115 & \textbf{0.0345} & 0.461 & 0.251 & \textbf{0.203} \\
 & LJ & 0.00833 & 0.0218 & 0.0171 & 0.0237 & \textbf{0.00735} & 0.0935 & 0.0456 & \textbf{0.0385} \\
\midrule
\multirow{7}{*}{\rotatebox[origin=c]{90}{Web}} & CW & 2.33 & 2.76 & 4.87 & 5.30 & \textbf{1.15} & 16.5 & 9.54 & \textbf{9.32} \\
 & SD & 0.260 & 0.412 & 0.348 & 0.926 & \textbf{0.174} & 1.60 & \textbf{1.07} & 1.07 \\
 & AR & 0.255 & 0.107 & 0.136 & 0.103 & \textbf{0.0507} & 0.368 & \textbf{0.169} & 0.173 \\
 & UK & 0.0437 & 0.0526 & 0.0895 & 0.0631 & \textbf{0.0249} & 0.300 & 0.125 & \textbf{0.113} \\
 & IC & 0.0631 & 0.0390 & 0.0620 & 0.102 & \textbf{0.0182} & 0.168 & 0.0772 & \textbf{0.0768} \\
\midrule
\multirow{7}{*}{\rotatebox[origin=c]{90}{Road}} & PL & 7.43 & 3.87 & 5.35 & 7.48 & \textbf{0.858} & 93.7 & 16.6 & \textbf{12.1} \\
 & EU & 3.03 & 1.56 & 2.77 & 3.12 & \textbf{0.420} & 44.5 & 7.17 & \textbf{5.00} \\
 & AS & 4.62 & 1.11 & 2.61 & 4.93 & \textbf{0.316} & 41.1 & 8.53 & \textbf{6.61} \\
 & AF & 2.01 & 0.251 & 1.08 & 2.31 & \textbf{0.109} & 16.5 & 3.18 & \textbf{2.57} \\
 & RU & 2.25 & 0.207 & 1.38 & 3.65 & \textbf{0.0624} & 6.08 & 4.66 & \textbf{4.55} \\
\midrule
\multirow{5}{*}{\rotatebox[origin=c]{90}{\shortstack{$k$-NN \&\\[-2pt] Synthetic}}} & COS & 1.64 & 3.45 & 4.45 & 1.48 & \textbf{1.04} & 10.0 & 6.00 & \textbf{4.93} \\
 & GL & 0.381 & 0.200 & 1.10 & 3.07 & \textbf{0.0745} & 2.45 & 1.20 & \textbf{1.08} \\
 & BBL & 2.55 & 0.181 & 1.23 & 3.50 & \textbf{0.0540} & 7.68 & 3.74 & \textbf{3.39} \\
 & TRCE & 1.78 & 0.135 & 1.08 & 2.62 & \textbf{0.0449} & 5.32 & 2.65 & \textbf{2.32} \\
 & CH & 0.230 & 0.287 & 1.27 & 5.00 & \textbf{0.160} & \textbf{0.599} & 1.91 & 1.61 \\
\bottomrule
\end{tabular}

\medskip

\begin{tabular}{llccccccccc}
\toprule
\multicolumn{2}{c}{} & \multicolumn{5}{c}{$k$-core} & \multicolumn{2}{c}{Bellman-Ford} & \multicolumn{2}{c}{Coloring} \\[2pt]
\cmidrule(lr){3-7}
\cmidrule(lr){8-9}
\cmidrule(lr){10-11}
 &  & GBBS & PASGAL & ParK & PKC & Ours & GBBS & Ours & GBBS & Ours \\
\midrule
\multirow{6}{*}{\rotatebox[origin=c]{90}{Social}} & FS & 5.77 & 3.14 & 3.36 & 5.22 & \textbf{2.67} & 3.06 & \textbf{2.48} & 5.38 & \textbf{4.01} \\
 & TW & 4.73 & 2.33 & 10.6 & 11.4 & \textbf{1.75} & 1.13 & \textbf{0.865} & 4.48 & \textbf{2.09} \\
 & OK & 1.17 & 0.435 & 0.515 & 0.504 & \textbf{0.265} & 0.111 & \textbf{0.0800} & 0.756 & \textbf{0.238} \\
 & FK & 0.331 & 0.387 & 0.321 & 0.251 & \textbf{0.0999} & 0.504 & \textbf{0.197} & 0.184 & \textbf{0.115} \\
 & LJ & 0.566 & 0.207 & 0.309 & 0.199 & \textbf{0.117} & 0.0801 & \textbf{0.0467} & 0.373 & \textbf{0.114} \\
\midrule
\multirow{7}{*}{\rotatebox[origin=c]{90}{Web}} & CW & 51.6 & 28.7 & 196 & 161 & \textbf{18.8} & 19.8 & \textbf{14.2} & 31.5 & \textbf{23.3} \\
 & SD & 6.16 & 3.61 & 24.7 & 9.78 & \textbf{3.12} & 2.54 & \textbf{1.87} & 7.53 & \textbf{4.10} \\
 & AR & 2.09 & 2.54 & 3.32 & 1.55 & \textbf{1.24} & 0.367 & \textbf{0.207} & 0.971 & \textbf{0.431} \\
 & UK & 0.958 & 0.823 & 1.01 & 0.422 & \textbf{0.368} & 0.271 & \textbf{0.115} & 0.385 & \textbf{0.166} \\
 & IC & 0.964 & 1.19 & 1.90 & 0.882 & \textbf{0.522} & 0.112 & \textbf{0.0669} & 1.14 & \textbf{0.411} \\
\midrule
\multirow{7}{*}{\rotatebox[origin=c]{90}{Road}} & PL & 2.75 & 1.95 & 2.23 & 1.10 & \textbf{0.961} & 122 & \textbf{76.2} & 3.00 & \textbf{1.91} \\
 & EU & 0.880 & 0.656 & 0.726 & 0.479 & \textbf{0.324} & 37.5 & \textbf{22.3} & 0.966 & \textbf{0.641} \\
 & AS & 0.674 & 0.469 & 0.520 & 0.404 & \textbf{0.234} & 31.9 & \textbf{19.7} & 0.694 & \textbf{0.466} \\
 & AF & 0.268 & 0.147 & 0.210 & 0.170 & \textbf{0.0866} & 8.35 & \textbf{4.44} & 0.222 & \textbf{0.165} \\
 & RU & 0.170 & 0.0984 & 0.157 & 0.121 & \textbf{0.0291} & 10.7 & \textbf{3.30} & 0.0928 & \textbf{0.0549} \\
\midrule
\multirow{5}{*}{\rotatebox[origin=c]{90}{\shortstack{$k$-NN \&\\[-2pt] Synthetic}}} & COS & 3.51 & 2.00 & 3.11 & 1.23 & \textbf{0.734} & 137 & \textbf{47.8} & 4.20 & \textbf{2.56} \\
 & GL & 0.296 & 0.150 & 0.214 & 0.162 & \textbf{0.0617} & 5.52 & \textbf{1.92} & 0.372 & \textbf{0.225} \\
 & BBL & 0.908 & 0.0739 & 0.148 & 0.0883 & \textbf{0.0422} & 17.6 & \textbf{7.80} & 0.121 & \textbf{0.0785} \\
 & TRCE & 1.30 & 0.0579 & 0.133 & 0.0610 & \textbf{0.0317} & 10.3 & \textbf{4.19} & 0.0847 & \textbf{0.0582} \\
 & CH & 0.0523 & 0.0223 & 0.0571 & \textbf{0.00619} & 0.00759 & 9.86 & \textbf{4.39} & 0.0352 & \textbf{0.0220} \\
\bottomrule
\end{tabular}
\medskip
\caption{\label{tab:rawdata_main}
Raw data corresponding to \cref{fig:main_fa}. Each entry is the arithmetic mean of 5 runs, in seconds.
For each problem and each graph instance, the fastest time is shown in bold. 
Baselines are introduced in \cref{sec:exp:app}.
IS: IndexSet; IM: IndexMap.
}
\end{table*}

\begin{table*}
\small
\renewcommand{\arraystretch}{0.9}
\begin{tabular}{ccc@{  }cc@{  }cc@{  }cc@{  }c}
\toprule
\multicolumn{2}{c}{} & \multicolumn{2}{c}{BFS} & \multicolumn{2}{c}{$k$-core} & \multicolumn{2}{c}{Coloring} & \multicolumn{2}{c}{Bellman-Ford} \\[2pt]
\cmidrule(lr){3-4}
\cmidrule(lr){5-6}
\cmidrule(lr){7-8}
\cmidrule(lr){9-10}
 &  & \hashbag & Ours & \hashbag & Ours & \hashbag & Ours & \hashbag & Ours \\
\midrule
\multirow{6}{*}{\rotatebox[origin=c]{90}{Social}} & FS & 0.260 & \textbf{0.102} & 3.02 & \textbf{2.74} & 4.15 & \textbf{4.01} & 3.07 & \textbf{2.48} \\
 & TW & 0.134 & \textbf{0.0452} & 4.43 & \textbf{4.10} & 2.18 & \textbf{2.09} & 1.19 & \textbf{0.865} \\
 & OK & 0.0187 & \textbf{0.00738} & 0.366 & \textbf{0.336} & 0.252 & \textbf{0.238} & 0.109 & \textbf{0.0800} \\
 & FK & 0.180 & \textbf{0.0341} & 0.357 & \textbf{0.105} & 0.328 & \textbf{0.115} & 0.900 & \textbf{0.197} \\
 & LJ & 0.0215 & \textbf{0.00845} & 0.204 & \textbf{0.156} & 0.163 & \textbf{0.114} & 0.0894 & \textbf{0.0467} \\
\midrule
\multirow{7}{*}{\rotatebox[origin=c]{90}{Web}} & CW & 5.11 & \textbf{1.16} & 66.6 & \textbf{61.2} & 27.1 & \textbf{23.3} & 40.3 & \textbf{14.2} \\
 & SD & 0.443 & \textbf{0.178} & 3.91 & \textbf{3.67} & 4.41 & \textbf{4.10} & 2.71 & \textbf{1.87} \\
 & AR & 0.309 & \textbf{0.0515} & 1.86 & \textbf{1.73} & 0.660 & \textbf{0.431} & 0.554 & \textbf{0.207} \\
 & UK & 0.108 & \textbf{0.0249} & 0.656 & \textbf{0.545} & 0.371 & \textbf{0.166} & 0.435 & \textbf{0.115} \\
 & IC & 0.113 & \textbf{0.0188} & 0.826 & \textbf{0.740} & 0.488 & \textbf{0.411} & 0.155 & \textbf{0.0669} \\
\midrule
\multirow{7}{*}{\rotatebox[origin=c]{90}{Road}} & PL & 3.68 & \textbf{3.07} & 2.24 & \textbf{1.19} & 3.37 & \textbf{1.91} & 101 & \textbf{76.2} \\
 & EU & 1.91 & \textbf{1.45} & 0.760 & \textbf{0.401} & 1.08 & \textbf{0.641} & 33.8 & \textbf{22.3} \\
 & AS & 2.01 & \textbf{1.41} & 0.563 & \textbf{0.307} & 0.782 & \textbf{0.466} & 28.4 & \textbf{19.7} \\
 & AF & 0.965 & \textbf{0.580} & 0.208 & \textbf{0.123} & 0.274 & \textbf{0.165} & 6.58 & \textbf{4.44} \\
 & RU & 1.38 & \textbf{0.484} & 0.148 & \textbf{0.0462} & 0.166 & \textbf{0.0549} & 11.7 & \textbf{3.30} \\
\midrule
\multirow{5}{*}{\rotatebox[origin=c]{90}{\shortstack{$k$-NN \&\\[-2pt] Synthetic}}} & COS & 2.61 & \textbf{1.77} & 2.15 & \textbf{0.890} & 3.74 & \textbf{2.56} & 95.3 & \textbf{47.8} \\
 & GL & 0.506 & \textbf{0.282} & 0.184 & \textbf{0.0832} & 0.311 & \textbf{0.225} & 5.14 & \textbf{1.92} \\
 & BBL & 1.45 & \textbf{0.681} & 0.604 & \textbf{0.254} & 0.163 & \textbf{0.0785} & 18.7 & \textbf{7.80} \\
 & TRCE & 1.01 & \textbf{0.471} & 0.505 & \textbf{0.206} & 0.118 & \textbf{0.0582} & 10.8 & \textbf{4.19} \\
 & CH & 0.632 & \textbf{0.353} & 0.0293 & \textbf{0.00893} & 0.0476 & \textbf{0.0220} & 10.8 & \textbf{4.39} \\
\bottomrule
\end{tabular}
\medskip
\caption{\label{tab:rawdata_hashbag}
Raw data for \cref{tab:main_table_hashbag} and \cref{fig:hashbag_comparison}. Each entry is the arithmetic mean of 5 runs, in seconds.
For each problem and each graph instance, the fastest time is shown in bold. 
\hashbag: this implementation uses the \hashbag data structure in PASGAL to replace Toggle Tree in our own implementations.
}
\end{table*}

%% file: experiments/main/hashbag_figures.tex
\begin{figure*}[!t]
  \centering
  \includegraphics[width=\linewidth]{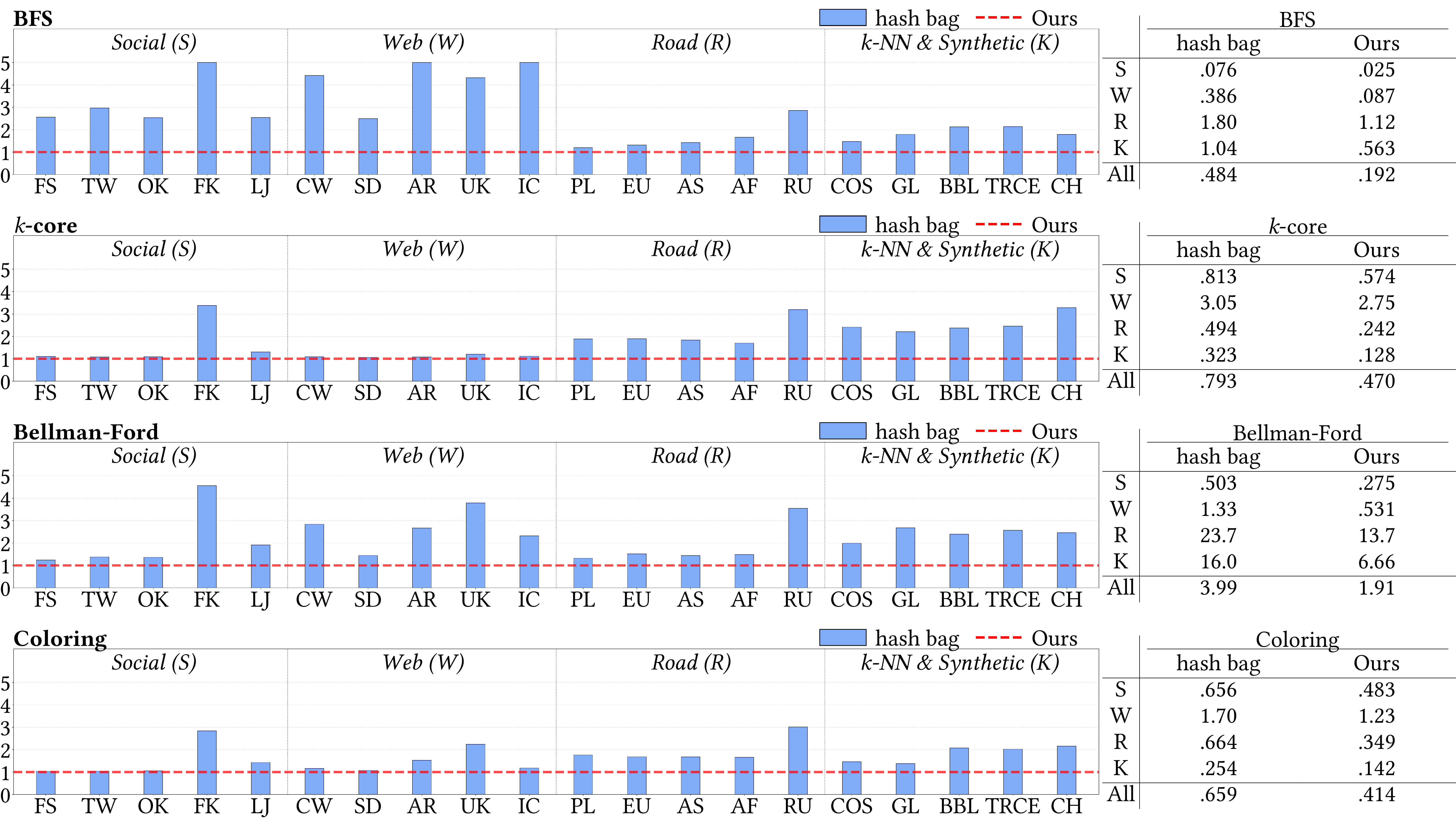}
  \caption{Comparison between \hashbag and our approach (Ours) for Bellman-Ford, BFS, coloring, and $k$-core.
    Blue bars show \hashbag running time normalized to ours.
    A higher bar means \hashbag is slower relative to ours.
    Values above 5 are clipped.
    }
  \label{fig:hashbag_comparison}
\end{figure*}